\documentclass[aps,prd,reprint,nofootinbib,floatfix,nofootinbib]{revtex4-2}

\usepackage[utf8]{inputenc} 

\usepackage{amsmath,amsfonts,amssymb}
\usepackage{bbold,bm}
\usepackage{slashed}

\usepackage{cancel}
\usepackage{graphicx}
\usepackage{orcidlink}
\usepackage[dvipsnames]{xcolor}

\usepackage{hyperref}
\hypersetup{colorlinks, linktocpage, urlcolor=BlueViolet, citecolor=Plum, linkcolor=PineGreen}

\usepackage{cleveref}

\DeclareMathOperator{\Tr}{Tr}
\DeclareMathOperator{\Diag}{diag}

\begin{document}

\title{Irreducible Constraints on Hadronically Interacting Sub-GeV Dark Matter}

\author{Peter Cox\,\,\orcidlink{0000-0002-6157-3430}}
\email{peter.cox@unimelb.edu.au}

\author{Matthew J. Dolan\,\,\orcidlink{0000-0003-3420-8718}}
\email{matthew.dolan@unimelb.edu.au}

\author{Avirup Ghosh\,\,\orcidlink{0000-0002-4781-842X}}
\email{avirup.ghosh@unimelb.edu.au}

\affiliation{ARC Centre of Excellence for Dark Matter Particle Physics, School of Physics, The University of Melbourne, Victoria 3010, Australia}

\begin{abstract}
We derive conservative upper limits on the dark-matter--nucleon scattering cross-section for sub-GeV mass dark matter. Working exclusively within the low-energy chiral effective theory, we derive bounds that are independent of the details of the dark matter interactions in the UV.  Dark matter that interacts only hadronically at leading order also inevitably interacts with photons or electrons at next-to-leading-order. We show that these electromagnetic interactions lead to strong constraints from big bang nucleosynthesis and over-production of dark matter via freeze-in at low temperatures, while the leading-order hadronic couplings face stringent constraints from meson decays. Combining these constraints, we rule out both spin-independent and spin-dependent dark-matter--nucleon scattering cross-sections $\gtrsim 10^{-36}\,{\rm cm}^2$ for dark matter masses in the keV - 100 MeV range. These bounds are several orders of magnitude stronger than the existing constraints from astrophysics and cosmology and have significant implications for future low-mass direct detection experiments.   
\end{abstract}

\maketitle

\section{Introduction}
\label{sec:intro}

Particle dark matter (DM) with a mass below the GeV scale has been the subject of much theoretical and experimental investigation in the past decade. In particular, sub-GeV dark matter which interacts with the Standard Model (SM) primarily via interactions with the nucleus has spurred an array of new experimental proposals and detection methods~\cite{Essig:2022dfa}. Although much of this work has occurred in the context of direct detection, this part of dark matter parameter space is also subject to constraints from flavour physics, colliders, cosmology, and astrophysics. It is important to have a comprehensive understanding of these, since they determine the thresholds and exposures required for direct detection experiments to probe new territory. 

The aim of this work is to bridge the gap between existing constraints on the non-relativistic DM-nucleon cross-section and the typically much stronger, but model-dependent, constraints that can be obtained from high-energy processes. More specifically, we aim to produce a set of robust upper bounds on the DM-nucleon cross-section, while making minimal assumptions about the form of the dark matter interactions at higher energies. We therefore restrict our focus to processes involving energies below a few hundred MeV. While in any given UV-complete model there may be stronger constraints, the ones we present below are ineluctable.

Constraints on hadronic dark matter couplings are usually plotted in the benchmark ``direct detection" plane, spanned by the dark matter mass $m_\chi$ and the dark matter-nucleon scattering cross-section, $\sigma^{\chi N}$. Given the non-relativistic nature of dark matter in the Milky Way halo, direct detection experiments involve momentum transfers of $\mathcal{O}(10^{-3} m_{\chi})$. On the other hand, constraints from cosmic ray up-scattering and particle colliders such as the Large Hadron Collider (LHC) require knowledge of the DM-SM interaction and particle content at much higher energies. Consequently, they are inherently more model dependent. 

Astrophysical and cosmological constraints are also derived at low energy and apply directly to $\sigma^{\chi p}$. This includes constraints from Milky Way sub-halos~\cite{Nadler:2019zrb,DES:2020fxi,Maamari:2020aqz,Buen-Abad:2021mvc}, Lyman-$\alpha$~\cite{Dvorkin:2013cea,Xu:2018efh,Rogers:2021byl}, cosmic microwave background (CMB) anisotropies~\cite{Chen:2002yh,Gluscevic:2017ywp,Slatyer:2018aqg}, UV luminosity functions~\cite{Das:2025bnr}, and galaxy cluster counts~\cite{He:2025npy}.  However, these provide only weak constraints on parameter space, typically bounding $\sigma^{\chi p} \lesssim \mathcal{O}(10^{-29}\,\textrm{cm}^{2})$ (for cross-sections which are independent of velocity). Our constraints on the cross-section will be orders of magnitude stronger than such bounds.

It is well known that the existence of a light mediator implies much stronger constraints on $\sigma^{\chi p}$~\cite{Dolan:2014ska,Krnjaic:2015mbs,Knapen:2017xzo,Bondarenko:2019vrb,Gori:2025jzu}. These constraints are derived from stellar emission, supernova 1987A, and on-shell mediator production in meson decays. Inevitably, these are more model dependent than the low-energy astrophysical constraints which can be obtained directly on the DM-proton cross-section, due to the freedom of choice in couplings and mediator mass.

In this work we take a complementary approach and derive conservative bounds that are insensitive to the specific details of the particle(s) mediating the DM-SM interaction. Depending on the specific observable, our constraints apply whenever the mediator mass(es) are greater than 10\,-\,500\,MeV. 

In previous work~\cite{Cox:2024rew}, two of the present authors demonstrated how constraints from meson decays, an irreducible DM abundance derived from freeze-in, and the existence of extra relativistic degrees-of-freedom (d.o.f.) during Big Bang Nucleosynthesis (BBN) can be derived directly using the $SU(3)$ chiral effective field theory. The resulting bounds, for both scalar and fermionic dark matter, were orders of magnitude stronger than the existing cosmological and astrophysical bounds on the dark matter-nucleon cross-section.

In this work, we extend the analysis of~\cite{Cox:2024rew} to incorporate CP-conserving vector, axial-vector and pseudoscalar operators. The vector operator leads to a spin-independent direct detection cross-section, while the axial vector and pseudoscalar lead to spin-dependent cross-sections (see also the recent article~\cite{Gori:2025jzu}).

We find results similar in spirit to~\cite{Cox:2024rew}: rare meson decays, dark matter overproduction via freeze-in, and extra relativistic d.o.f. during BBN lead to very strong constraints on the relevant direct detection cross-sections. For dark matter masses below about $100\,$MeV, direct detection experiments need to be sensitive to cross-sections around $\mathcal{O}(10^{-36})~\textrm{cm}^2$ to probe new parameter space.

In next section, we codify the assumptions we make and observables we use. In sections~\ref{sec:vector}, \ref{sec:axialvector} and \ref{sec:pseudoscalar} we deal in turn with vector, axial-vector and pseudoscalar operators. Section~\ref{sec:gluon} discusses the topological gluon operator, before we present conclusions in \cref{sec:conclusion}. Some technical calculations are presented in the \cref{app:EFT,app:thermalisation,app:flavour}.

\section{Framework and Observables}
\label{sec:framework}

Our starting point involves the following set of assumptions:
\begin{enumerate}
    \item DM interacts only hadronically at leading order.\footnote{Also allowing for leading-order interactions with photons or leptons generically leads to stronger constraints.}
    \item The mediator(s) between the DM and SM have mass greater than\footnote{Some of the bounds we derive, such as from BBN and freeze-in, apply even for lower mediator masses $\gtrsim10$\,MeV.} $\mathcal{O}(100)$\,MeV.
    \item DM interacts with the SM via an operator that is bilinear in the DM field $\chi$. (Motivated by a $U(1)$ or $\mathbb{Z}_2$ stabilising symmetry.)
    \item The DM field $\chi$ is a singlet under $SU(3)_c \times U(1)_Q$.
\end{enumerate}
Schematically, we consider DM interactions of the form
\begin{equation}
    \mathcal{L} \supset \frac{c_{\alpha\beta}}{\Lambda^n} \mathcal{O}^\alpha_{\rm{SM}} \mathcal{O}^\beta_\chi \,,
    \label{eq:Lag_OO}
\end{equation}
where $\mathcal{O}^\alpha_{\rm{SM}}$ are local, gauge-singlet operators consisting of quark and/or gluon fields and the operators $\mathcal{O}^\beta_\chi$ are bilinear in $\chi$ and can be non-local. The $c_{\alpha\beta}$ are dimensionless coefficients and $\Lambda$ parametrises the scale of the UV physics, with $n=\text{dim}(\mathcal{O}_\text{SM})+\text{dim}(\mathcal{O}_\chi)-4$. We follow the same philosophy as our earlier work~\cite{Cox:2024rew}, which considered the case where $\mathcal{O}_\chi$ and $\mathcal{O}_\text{SM}$ are Lorentz scalars, but here extend the analysis to more general Lorentz structures.

For $\mathcal{O}_\text{SM}$, we consider the lowest dimension (dimension 3 and 4) quark and gluon operators listed in the left column of \cref{tab:operators}. We omit the tensor quark operator, since its coefficient is suppressed in most UV completions. We also restrict to the case where the Lagrangian in \cref{eq:Lag_OO} preserves $CP$, leaving the $CP$-violating case for future work.

\subsection{\texorpdfstring{$SU(3)$ Chiral EFT}{SU(3) Chiral EFT}}
\label{sec:SU3ChPT}

Since we are only interested in low-energy processes, we work entirely within the $SU(3)$ chiral effective theory where the DM has local interactions with mesons and baryons. The matching of the Lagrangian \eqref{eq:Lag_OO} (given in terms of quark and gluon fields) onto the Chiral Lagrangian can be performed by treating the DM operators as external sources that transform under the $SU(3)_L \times SU(3)_R$ symmetry~\cite{Gasser:1984gg,Bishara:2016hek}. (See \cref{app:EFT} for details.) 

We consider fermionic and scalar DM and assume that the mediator(s) between the DM and SM sectors is sufficiently heavy that the interactions in the low-energy chiral EFT are local and contain the lowest dimension operators bilinear in $\chi$ with the appropriate Lorentz and $CP$ properties. The operators we consider are summarised in \cref{tab:operators}.

For Dirac fermion DM, the matching of the quark operators onto the chiral EFT leads to the following (CP-preserving) interactions between the DM and the octet of pseudoscalar mesons at leading order in $1/\Lambda$ and up to $\mathcal{O}(p^2)$ in the chiral power counting,
\begin{align} \label{eq:chiralEFT-quark}
    \mathcal{L}_q = &-\frac{if^2}{2\Lambda^2} (\bar\chi\gamma^\mu\chi) \Big( \Tr[ C^V ( U \partial_\mu U^\dagger + U^\dagger \partial_\mu U)] \notag \\
    & + 2ie A_\mu \Tr[ C^V (U Q_q U^\dagger + U^\dagger Q_q U ) ] \Big) \notag \\
    &- \frac{if^2}{2\Lambda^2} (\bar\chi\gamma^\mu\gamma^5\chi) \Tr[C^A ( U \partial_\mu U^\dagger - U^\dagger \partial_\mu U) ] \notag \\
    & - \frac{if^2B_0}{2\Lambda^2} (\bar\chi i\gamma^5 \chi) \Tr[C^P (U-U^\dagger) ] \notag \\
    &  + \frac{f^2B_0}{2\Lambda^2} (\bar\chi \chi) \Tr[C^S (U+U^\dagger) ] \,.
\end{align}
Here, the Wilson coefficients $C^V, C^A, C^P, C^S$ are $3 \times 3$ matrices that parameterise the flavour structure of the interactions. The meson fields are encoded in $U=\exp{(i\sqrt{2}\Pi/f)}$, with $\Pi$ the hermitian matrix of mesons and $f\approx92$\,MeV the pion decay constant at leading order. Lastly, $A_\mu$ is the photon field, $e$ the electromagnetic coupling, $Q_q=\Diag(2/3,-1/3,-1/3)$ the matrix of the quark charges, and $B_0=2.6$\,GeV (see e.g.~\cite{Bishara:2016hek}) is a low-energy constant.

One could also consider the cases where the vector DM current couples to the axial-vector quark current and vice versa (i.e. interchanging the DM currents between the first and third lines of \cref{eq:chiralEFT-quark}). The constraints on these scenarios will be similar to those we derive for the purely vector and axial-vector interactions; however, an important difference that is the mixed vector/axial-vector interactions lead to velocity or momentum suppressed DM-nucleon scattering in the non-relativistic limit~\cite{Kumar:2013iva} and so are even more challenging to probe via direct detection. For this reason we do not consider them in detail.

To obtain the Lagrangian for complex scalar DM, one makes the following replacements in \cref{eq:chiralEFT-quark}:
\begin{align}
    \bar\chi\chi &\to \chi^*\chi \,, \\
    \bar\chi\gamma^\mu\chi &\to i\chi^* \overset{\leftrightarrow}{\partial^\mu}\chi \,,
\end{align}
where $\chi^* \overset{\leftrightarrow}{\partial^\mu}\chi \equiv \chi^*(\partial^\mu\chi) - (\partial^\mu\chi^*)\chi$. Note that for scalar DM there are no $CP$-conserving axial-vector or pseudoscalar interactions. While we explicitly consider Dirac fermion and complex scalar DM, out results can be straightforwardly re-interpreted for Majorana or real scalar DM, noting that the vector interactions vanish for these cases.

For the gluon operators, the matching onto the chiral EFT at $\mathcal{O}(p^2)$ yields
\begin{align} \label{eq:chiralEFT-gluon}
     \mathcal{L}_G = & \frac{f^2 C^G}{18\Lambda^3} (\bar\chi\chi) \left( \Tr[\partial^\mu U^\dagger \partial_\mu U] + 3 B_0 \Tr[M_q (U + U^\dagger )] \right) \notag \\
     & -\frac{i f^2 C^{\tilde G}}{12 \Lambda^3} \partial^\mu (\bar\chi i \gamma^5 \chi) \Tr[( U \partial_\mu U^\dagger - U^\dagger \partial_\mu U) ] \notag\\
     & -  \frac{i B_0 f^2 C^{\tilde{G}}}{6 \Lambda^3} (\bar\chi i \gamma^5 \chi) \Tr[M_q\,(U-U^\dagger)] \,,
\end{align}
with Wilson coefficients $C^G$ and $C^{\tilde{G}}$. Note that $C^{\tilde{G}}$ can be re-expressed in terms of $C^A$ and $C^P$. However, for convenience, we also provide constraints on the coefficient $C^{\tilde{G}}$ since the $G\tilde{G}$ operator commonly arises in minimal UV completions with a pseudoscalar mediator.

The DM interactions with protons and neutrons can similarly be obtained by matching onto heavy baryon chiral EFT; we refer the reader to Ref.~\cite{Bishara:2016hek} for the details.

In this paper, we focus on the vector, axial-vector, and pseudoscalar interactions ($C^V$, $C^A$, $C^P$, $C^{\tilde{G}}$) and refer the reader to our earlier work~\cite{Cox:2024rew} for constraints on the scalar interactions ($C^S$, $C^G$). The Wilson coefficients $C^{V,A,S,P}$ have, in general, non-trivial flavour structure. Here, we focus on flavour-diagonal couplings and present results for the following cases:
\begin{equation}
    C^V = \mathbb{1}, Q_q \,, \qquad C^A = \mathbb{1}, Q_q \,, \qquad C^P = M_q /\Lambda \,,
\end{equation}
where $M_q=\Diag(m_u,m_d,m_s)$. These choices are motivated by minimal UV completions, as we discuss in  \cref{sec:vector,sec:axialvector,sec:pseudoscalar}. While we present results only for these particular flavour structures, we stress that other choices are expected to be more \emph{strongly constrained}. There are two reasons for this. First, with flavour-universal couplings $C^{V,A,P} \propto \mathbb{1}$, the leading order interaction terms vanish. The flavour-universal scenarios are therefore expected to be the least constrained for the vector and axial-vector operators. (The flavour-universal pseudoscalar case is more subtle, as we discuss in \cref{sec:pseudoscalarmodels}). For these cases, it is necessary to use the next-to-leading-order (NLO) Chiral Lagrangian; we discuss this in detail in the relevant sections. Second, flavour off-diagonal couplings are strongly constrained by flavour-changing-neutral-current (FCNC) meson decays.

\begin{table}[t]
    \centering
    \begin{tabular}{|c||c|c|}
    \hline
    $\mathcal{O}_{\rm SM}$ & $\mathcal{O}_{\chi}$ (scalar DM) & $\mathcal{O}_{\chi}$ (fermionic DM) \\ 
    \hline
    $\bar{q_i} q_j$ \cite{Cox:2024rew} & $\chi^* \chi$ & $\bar{\chi}\chi$ \\ 
    $\bar{q_i}\gamma_\mu q_j$ & $i\chi^* \overset{\leftrightarrow}{\partial^\mu}\chi$ & $\bar{\chi}\gamma^\mu\chi$ \\ 
    $\bar{q_i}\gamma_\mu \gamma_5 q_j$ & - & $\bar{\chi}\gamma^\mu\gamma^5 \chi$ \\
    $i\bar{q_i}\gamma_5 q_j$ & - & $\bar{\chi}i\gamma^5\chi$ \\
    \hline
    $\frac{\alpha_s}{8\pi} G_{a\mu\nu}G^{a\mu\nu}$ ~\cite{Cox:2024rew} & $\chi^* \chi$ & $\bar{\chi}\chi$ \\ 
    $\frac{\alpha_s}{8\pi} G_{a\mu\nu}\tilde{G}^{a\mu\nu}$ & - & $\bar{\chi}i\gamma^5\chi$ \\
    \hline
    \end{tabular}
    \caption{SM and DM bilinear operators considered in this work. Note that for scalar DM there are no $CP$-preserving axial-vector or pseudoscalar interactions.} 
    \label{tab:operators}
\end{table}

\subsection{Observables/Constraints}
\label{sec:obscons}

We consider only low-energy observables that can be computed within the chiral EFT introduced in the previous section. This allows us to avoid making assumptions about the form of the DM interactions at higher energies. As we shall show, these observables are nevertheless sufficient to derive stringent constraints on the DM interactions. Rather than providing bounds on the EFT scale $\Lambda$, we re-express our results as constraints on the spin-independent/spin-dependent DM-nucleon cross-section. This allows for a direct comparison with existing astrophysical and cosmological bounds and makes manifest the implications for low-mass direct detection experiments.

\subsubsection{Big Bang Nucleosynthesis}
\label{sec:bbn}

The success of SM BBN tightly constraints the abundances of any additional relativistic degrees of freedom during neutrino decoupling and nucleosynthesis. In fact, sub-MeV mass states that were in equilibrium at any time after the QCD phase transition but before the end of BBN are essentially excluded. More specifically, an electromagnetically interacting Dirac fermion (complex scalar) with $m_\chi<0.7\ (0.5)$\,MeV is excluded if it was in equilibrium during neutrino coupling and BBN~\cite{Boehm:2013jpa,Depta:2019lbe,Sabti:2019mhn,Sabti:2021reh}. This remains true even if these states decouple from the SM during BBN~\cite{Cox:2024rew}.

Sub-MeV mass DM that interacts \emph{only} hadronically could, however, easily remain out of equilibrium during BBN, since the abundances of the mesons and baryons are exponentially suppressed at the relevant temperatures. However, DM that interacts hadronically \emph{inevitably} also couples to photons and electrons. Within the chiral EFT these couplings arise at $\mathcal{O}(p^4)$ in the chiral counting, for example, via one-loop diagrams involving charged pions and kaons (see~\cref{fig:EEChiChi}) or (either directly or indirectly) through the Wess-Zumino-Witten (WZW) term. While these interactions are suppressed compared to the leading order hadronic interactions, they may still be sufficient to bring the DM into equilibrium with the SM. 

As well as remaining agnostic to the details of the DM interactions at high energies, we want to avoid making unnecessary assumptions about the early cosmological history of the Universe. While the reheat temperature of the Universe may have been very high, viable BBN only requires that the SM plasma once reached a temperature of around $5 - 10$\,MeV~\cite{Hannestad:2004px,Barbieri:2025moq}. The constraint we impose is therefore to require that the DM was out of equilibrium at a temperature of $T=10\,$MeV. Stronger bounds can be derived if a higher reheat temperature is assumed.

\subsubsection{Irreducible DM abundance}
\label{sec:DMFI}

We do not require that $\chi$ saturates the DM relic abundance (except when showing existing bounds from direct detection). Once again, this allows us to remain insensitive to the early cosmological history and the specifics of the DM production mechanism. We do, however, require that any DM produced at low temperatures (below 10\,MeV) does not exceed the observed relic abundance. 

Even if the DM remains out of equilibrium with the SM at temperatures below 10\,MeV (thus satisfying the constraint from BBN), there is an \emph{irreducible} freeze-in abundance of DM produced at low temperatures. Since the meson and baryon abundances are suppressed, this is produced predominantly through the electromagnetic interactions discussed above, via processes such as $\gamma\gamma \to \chi\bar\chi$ and $e^+ e^- \to \chi\bar\chi$. 

If this freeze-in abundance exceeds the DM abundance as measured by CMB observations, it is possible, in principle,  for it to be depleted prior to recombination. This could happen through annihilations, decay, or dilution due to entropy production, depending on the details of the specific model. However, in practice, significantly depleting an excess of DM through any of these mechanisms is very difficult without either violating the CMB bound on dark radiation ($\Delta N_\text{eff}$) and/or diluting the baryons.

\subsubsection{Meson decays}
\label{sec:mesondecay}

Meson decays can provide strong constraints on the hadronic interactions of DM. In particular, the decay rates of $\pi$, $K$, and $B$ mesons to final states with missing energy are all tightly constrained. This is especially true for the FCNC decays $K \to \pi + \text{invisible}$, $B \to K + \text{invisible}$, and $B \to \pi + \text{invisible}$. 

In this work, we do not consider bounds from invisible decays of the heavier $B$ (or $D$) mesons. While this might seem overly conservative, it means that we do not have to specify the Lagrangian describing the DM interactions at GeV energies and can work solely within the chiral EFT.

The decay modes of relevance and the precise bounds on the DM interactions depend on both the type of SM operator and the flavour structure of the Wilson coefficients (for the quark operators). For a particular choice of the Wilson coefficients it is therefore be possible to suppress a given decay mode. However, it is important to stress that while the constraints from a particular meson decay might be evaded, it is generally not possible to simultaneously suppress all of the meson decay channels. We discuss an explicit example of this below.

For the vector, axial-vector, and scalar interactions ($C^V$,$C^A$,$C^S$,$C^G$), the decay $K^+ \to \pi^+ \chi\bar\chi$ provides the dominant constraint in most cases\footnote{Note that the decay $K^0_L \to \pi^0 \chi\bar\chi$, which is constrained by the KOTO experiment~\cite{KOTO:2024zbl}, is $CP$ violating for all of the operators we consider and is therefore suppressed.}, with the NA62 experiment constraining the branching ratio to be $\lesssim 2\times10^{-10}$~\cite{NA62:2024pjp}. If the Wilson coefficients are flavour-diagonal, then the $s\to d$ transition is mediated by the SM weak interactions. In the chiral EFT, the leading order $\Delta S = 1$ Lagrangian is given by~\cite{Gerard:2005yk}
\begin{multline} \label{eq:DeltaS1Lagterm1}
    \mathcal{L}^{\rm LO}_{\Delta S=1} \supset \frac{G_F}{\sqrt{2}}V_{ud}V^*_{us} f^4 \Big( g_8 \Tr[ \lambda L_\mu L^\mu] 
    + g^S_8 (L_\mu)_{32} \Tr[L^\mu] \\ 
    + g_{27} \big[ (L_\mu)_{32}(L^\mu)_{11} + \frac{2}{3}(L_\mu)_{12} (L^\mu)_{31} - \frac{1}{3}  (L_\mu)_{32} \Tr[L^\mu] \big] \Big) + {\rm h.c.} \,,
\end{multline}
where $L^\mu = U^\dagger \mathcal{D}^\mu U$, $\lambda = (\lambda_6-i \lambda_7)/2$ with $\lambda_{6,7}$ the Gell-Mann matrices, $G_F$ is the Fermi constant, $V_{ij}$ are CKM matrix elements, and $g_8=3.07\pm0.14$~\cite{Pich:2021yll}, $g^S_8 \approx -1.17$~\cite{Gerard:2005yk}, and $g_{27}=0.29 \pm 0.02~\cite{Pich:2021yll}$ are low-energy constants. The covariant derivative $\mathcal{D}_\mu U$, defined in \cref{eq:covDeriv}, also includes vector and axial-vector interactions with the DM, as discussed in \cref{app:EFT}. Note that if the Wilson coefficients ($C^{V,A,S}$) have non-zero off-diagonal entries then there can be additional $\Delta S=1$ terms in the low-energy EFT, which are not suppressed by $G_F$; these generally lead to much stronger constraints from $K^+ \to \pi^+ \chi\bar\chi$.

For the two representative flavour structures we consider for the vector interactions ($C^V\propto \mathbb{1},Q_q$), it turns out that the $K^+ \to \pi^+ \chi\bar\chi$ amplitude vanishes. We discuss this in detail in \cref{sec:vector}. The leading meson decay constraint then comes from $\pi^0 \to \gamma \chi\bar\chi$, which is mediated by the WZW interaction. The branching ratio of this decay mode is constrained to be $< 1.9\times10^{-7}$~\cite{NA62:2019meo}.

The pseudoscalar interactions ($C^P$ and $C^{\tilde{G}}$) are qualitatively different, since the $K \to \pi \chi\bar\chi$ decay is mediated via mixing with the neutral pion. The dominant constraint then comes from the invisible decay of the on-shell pion $\pi \to \chi\bar\chi$, with an upper limit on the branching ratio of $4.4\times10^{-9}$~\cite{NA62:2020pwi}. For heavier DM masses, the decay $\eta \to \chi\bar\chi$ becomes relevant, although with a much weaker bound of $1.1\times10^{-4}$ on the branching ratio~\cite{NA64:2024mah}.

\subsubsection{Other constraints}

There are many other observables that can be used to constrain the DM parameter space, but which we do not consider here. First, there are those observables that require the specification of a Lagrangian valid at higher energies, beyond a few hundred MeV. These include searches for missing energy at the Large Hadron Collider, direct detection of cosmic-ray up-scattered DM, and flavour constraints from heavier mesons (such as $B$-decays). In any given UV-complete model, these may well be more constraining than what we present here. 

Second, there are observables that also require the specification of a DM production mechanism and complete cosmological history. Of particular note is the bound on warm DM from structure formation as probed by Lyman-$\alpha$. If the DM was produced via freeze-in, for example, this sets a lower bound on the DM mass, typically in the $\mathcal{O}(1)-\mathcal{O}(10)$\,keV range~\cite{Ballesteros:2020adh}. There is also the CMB constraint on the effective number of relativistic degrees of freedom during recombination ($\Delta N_\text{eff}$); however, this is correlated with the DM abundance and the precise bound is therefore dependent on the DM production mechanism.

Last, there are energy loss bounds on the emission of new light particles from stars and SN1987A. These set stringent limits on the couplings of light scalar, pseudoscalar, and vector mediators with masses below 100\,MeV (see e.g.~\cite{Hardy:2016kme,Lella:2023bfb,Bottaro:2023gep,Hardy:2024gwy,Fiorillo:2025zzx}). For heavier mediator masses, as considered here, energy loss arguments can constrain the pair production and emission of light DM; however, to the best of our knowledge, precise bounds are yet to be calculated. Nevertheless, a na\"ive estimate suggests that these will not apply for cross-sections $\sigma^{\chi N} \gtrsim 10^{-45}\,\text{cm}^2$ ($\sigma^{\chi N} \gtrsim 10^{-38}\,\text{cm}^2$), since the DM would become trapped in the SN (star). As such, stellar/SN bounds are expected to be highly complementary to those we present in this work.

\section{Vector Operator}
\label{sec:vector}

We begin with the vector operator (second row of \cref{tab:operators}). The leading order DM-meson interactions are given by the $C^V$-dependent terms in the first two lines of \cref{eq:chiralEFT-quark}. We also include the Lagrangian terms that contribute to the DM interactions with photons and electrons at $\mathcal{O}(p^4)$ in the chiral counting. These are the $L_{10}$ and $H_1$ terms and the Wess-Zumino-Witten (WZW) functional (see e.g.~\cite{Pich:1995bw}). Expanding these terms in the meson fields gives
\begin{align}
    \mathcal{L} &= \,\frac{i}{\Lambda^2}(\bar{\chi}\gamma^\mu \chi)\,
    \Big[ (C^V_u-C^V_d) (\pi^- \overset{\leftrightarrow}{\partial}_\mu \pi^+ + 2i e A_\mu \pi^- \pi^+) \nonumber \\
    & + (C^V_u-C^V_s) (K^- \overset{\leftrightarrow}{\partial}_\mu K^+ + 2 i e A_\mu K^- K^+)\Big]  \nonumber \\ 
    &- \frac{4e}{3\Lambda^2} \left( L_{10} + 2H_1 \right) \left(2C^V_u-C^V_d-C^V_s \right) \partial^\mu \left( \bar{\chi}\gamma^\nu \chi\right) F_{\mu\nu} \nonumber \\
    &+\frac{e}{8\pi^2 f \Lambda^2}\left(2C^V_u+C^V_d\right) \,\pi^0\,\partial^\mu \left( \bar{\chi}\gamma^\nu \chi\right) F^{\rho\sigma}\epsilon_{\mu\nu\rho\sigma} + \ldots \,,
\label{eq:DMVcoupChPT}
\end{align}
where we have taken $C^V = \Diag(C^V_u, C^V_d, C^V_s)$. Here, $F^{\rho\sigma}$ is the electromagnetic field strength tensor and $\epsilon_{\mu\nu\rho\sigma}$ is the 4-dimensional Levi-Civita tensor. The low-energy constant $L_{10}=-3.8\times10^{-3}$~\cite{Bijnens:2014lea}. (We comment on the remaining low-energy constant $H_1$ below.)  The fourth line is the WZW contribution. The equivalent Lagrangian for complex scalar DM can be obtained by replacing $\bar\chi\gamma^\mu\chi \to i\chi^* \overset{\leftrightarrow}{\partial^\mu}\chi$. 

We will trade the scale $\Lambda$ in the Lagrangian above for $\sigma^{\chi N}_{\rm SI}$ when presenting our constraints to compare against existing results from direct detection experiments and cosmology.   
In the non-relativistic (NR) limit, the spin-independent scattering cross-section of DM with nucleons is given by (see e.g.~\cite{Cirelli:2013ufw})
\begin{equation}
    \sigma^{\chi N}_{\rm SI} = \frac{\mu^2_{\chi N}}{\pi \Lambda^4} \bigg(\sum_{q=u,d} N_q C^V_q \bigg)^2 \,, 
    \label{eq:SIDDvectorcoup}
\end{equation}
where $\mu_{\chi N} = m_\chi m_N/(m_\chi+m_N)$ is the reduced mass of the DM-nucleon system, and $N_q$ represents the number of valence quarks of flavour `$q$' within nucleon `$N$'. 

\begin{figure}[t]
    \centering
    \includegraphics[width=0.5\textwidth]{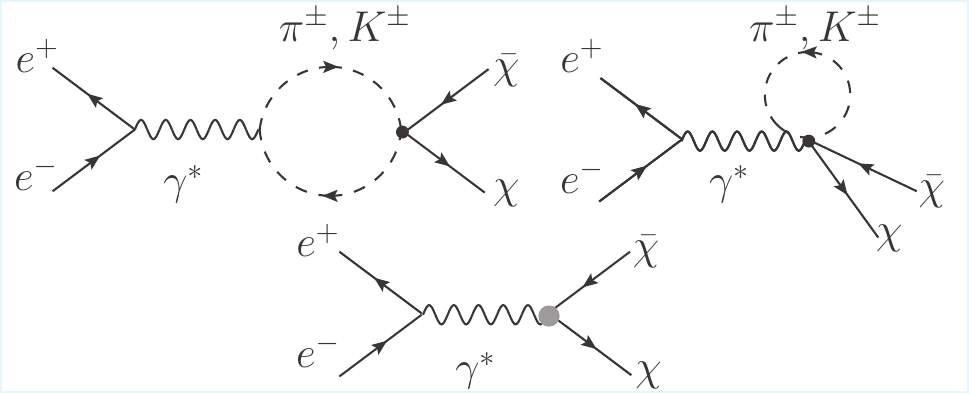}  
    \caption{Feynman diagrams that contribute to the DM thermalisation and freeze-in production process $e^+e^- \to \chi \bar\chi$ for DM interacting via the quark vector current.} 
    \label{fig:EEChiChi}
\end{figure}

\begin{figure}[t]
    \centering
    \includegraphics[width=0.9\linewidth]{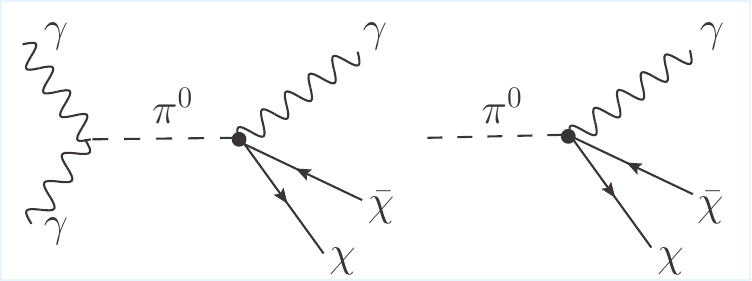}
    \caption{Feynman diagrams contributing to DM thermalisation and freeze-in production (left) and pion decay (right) for the vector operator via the WZW term.} 
    \label{fig:VcoupMeson}
\end{figure}

The relevant constraints are those discussed in \cref{sec:framework}. The $\gamma\gamma\to \chi\bar\chi$ process vanishes for the Lagrangian in \cref{eq:DMVcoupChPT} due to charge conjugation symmetry. Hence, at low temperatures, DM is primarily created by $e^+ e^- \to \chi \bar{\chi}$. This process occurs at $\mathcal{O}(p^4)$ in the chiral counting via the diagrams in \cref{fig:EEChiChi}. The tree-level diagram comes from the third line of \cref{eq:DMVcoupChPT} and is proportional to\footnote{Note that $H_1$ is un-measureable in the SM, but its counterterm is needed to renormalise the DM-photon interaction. In computing $e^+ e^- \to \chi \bar{\chi}$, we neglect the unknown finite part.} $(L_{10}+2H_{1})$. The interaction rate for this process is given in \cref{eq:Geechichi}. Parameter regions where this process is in equilibrium at $T=10$\,MeV are disallowed by BBN. For DM which remains out of thermal equilibrium, the irreducible freeze-in abundance is produced by the same process and has been calculated using \cref{eq:FIvectorcoup}. The WZW term gives rise to the process $\gamma\gamma \to \gamma\chi \bar{\chi}$ (see \cref{fig:VcoupMeson} left), which also contributes to DM production and thermalisation; however, this process is sub-dominant to $e^+ e^- \to \chi \bar{\chi}$ (due to the additional factor of $\alpha$ and the three-body phase space), except when $C^V=\mathbb{1}$.

For meson decays, we provide details of our calculation of $K^+ \to \pi^+ \chi \bar{\chi}$ in \cref{app:flavour}. The decay rate is given in \cref{eq:KaondecayVector} and reduces to
\begin{align}
    \Gamma_{K^+ \to \pi^+ \chi \bar{\chi}} \propto \,& \frac{\sigma^{\chi N}_{\rm SI}}{\mu^2_{\chi N}} G^2_F |V_{ud}|^2 |V^{*}_{us}|^2 f^4 \left( 3g_8 + 2g_{27}\right)^2  \notag\\
    & \left( C^V_d - C^V_s \right)^2 m^5_{K^+} \,,
    \label{eq:Kaondecayvector}
\end{align}
in the $m_{K^+} \gg m_{\pi^+}, m_\chi$ limit. Notice that this vanishes\footnote{However, we have not computed the $\mathcal{O}(p^4)$ corrections to this decay, which include terms proportional to the quark masses that may not vanish when $C^V_d = C^V_s$.} for $C^V_d = C^V_s$.

There is another meson decay constraint, which arises from the WZW interaction term in the fourth line of \cref{eq:DMVcoupChPT}. This leads to the decay $\pi^0 \to \gamma\chi \bar{\chi}$ (see \cref{fig:VcoupMeson} right), which is constrained by the upper-limit on the branching ratio of $\pi^0 \to \gamma\nu \bar{\nu}$ from NA62~\cite{NA62:2019meo}. The decay rate is given in \cref{eq:pi03bodydecays} and is proportional to $(2C^V_u+C^V_d)^2$.

\subsection{Models and Results}
\label{sec:vectormodels}

The quantitative bounds on $\sigma^{\chi N}$ depend on the flavour structure of the Wilson coefficient $C^V$. We present constraints for two specific cases, $C^V=\mathbb{1}$ and $C^V=Q_q$. We choose these scenarios because (i) the flavour-universal case is the \emph{least} constrained, and (ii) they arise from the minimal UV completions. 

Before presenting our results, let us briefly comment on UV completions. The minimal, tree-level completion involves a new $U(1)_X$ gauge symmetry under which the DM is charged. To obtain models with only hadronic couplings at leading order, the quarks also need to be charged under $U(1)_X$.\footnote{If the only coupling to the SM is via kinetic mixing, one arrives at the well-studied dark photon portal. The leading-order couplings to leptons lead both to additional constraints and the possibility of detecting the DM via electron scattering.} Such symmetries will, in general, be anomalous and additional chiral fermions are needed to cancel the anomaly. These can be made massive via a Yukawa coupling to a $U(1)_X$-breaking Higgs that also gives mass to the gauge boson. The simplest realisation is where the quarks have universal charges (i.e. $X$ = baryon number), which leads to $C^V=\mathbb{1}$ at low energies. The case $C^V=Q_q$ is obtained with $X=Y-L$, where $Y$ is hypercharge and $L$ lepton number. Other flavour-dependent cases require non-minimal extensions to the quark/Higgs sector to generate the quark masses and mixing. Other UV completions besides a $U(1)$ vector boson are of course also possible, but will often also generate other operators in the low energy EFT.

While we have sketched out some minimal UV completions above, we reiterate that the bounds we present below are independent of any specific UV model (which will be subject to additional model-dependent constraints).

\begin{figure*}[t]
    \centering
    \includegraphics[width=0.50\linewidth]{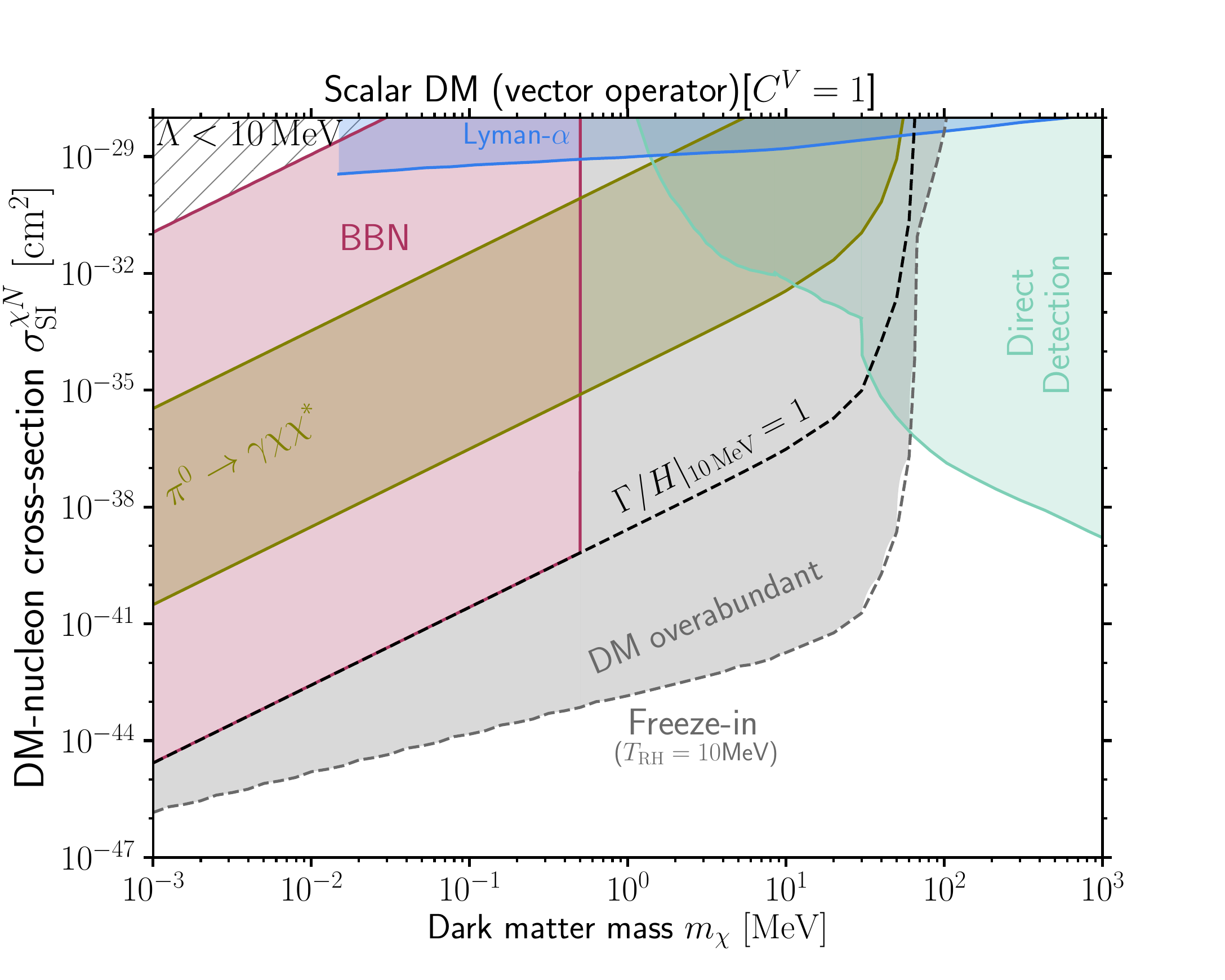}\!\!\!\!\!\!\!
    \includegraphics[width=0.50\linewidth]{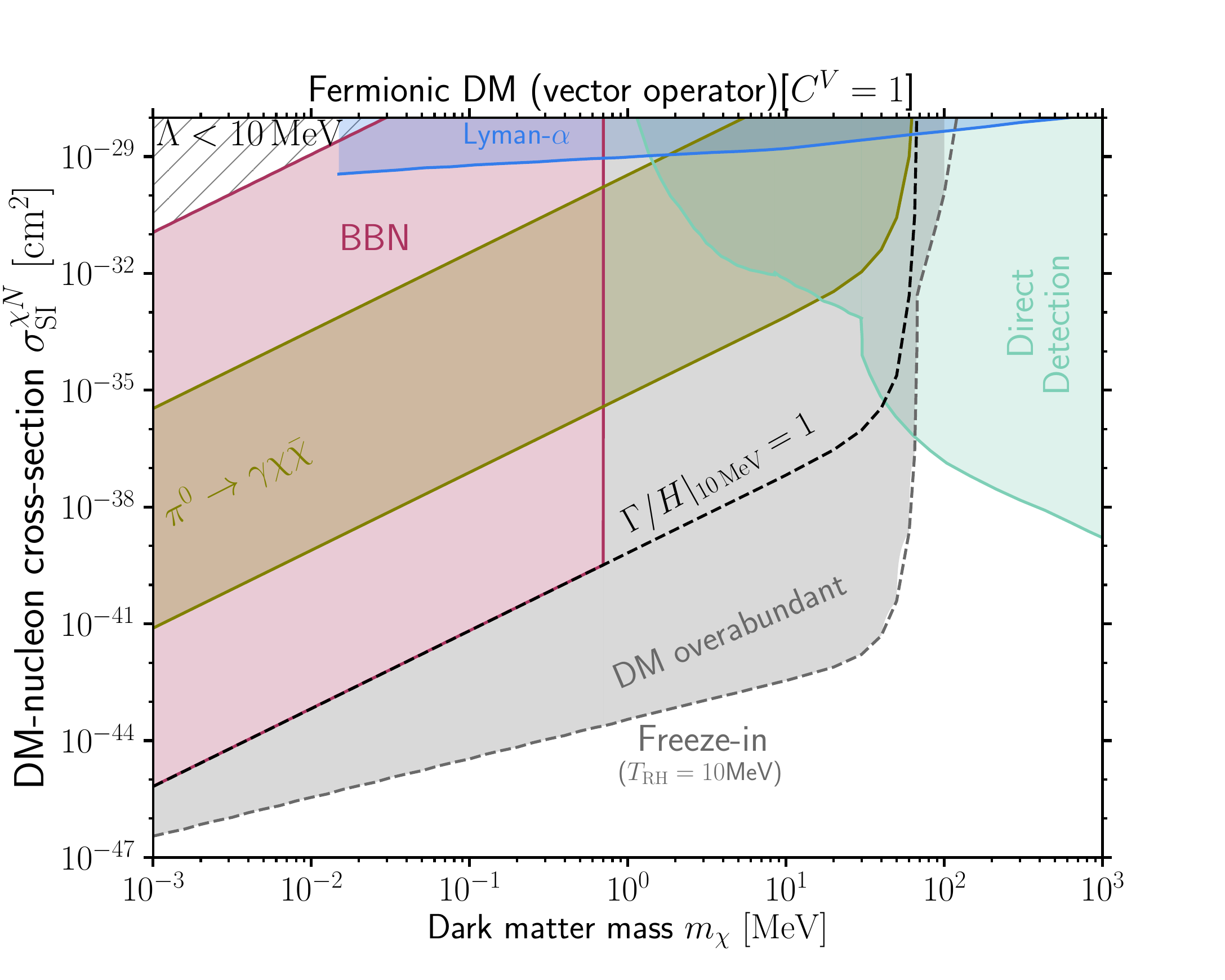}
    \includegraphics[width=0.50\linewidth]{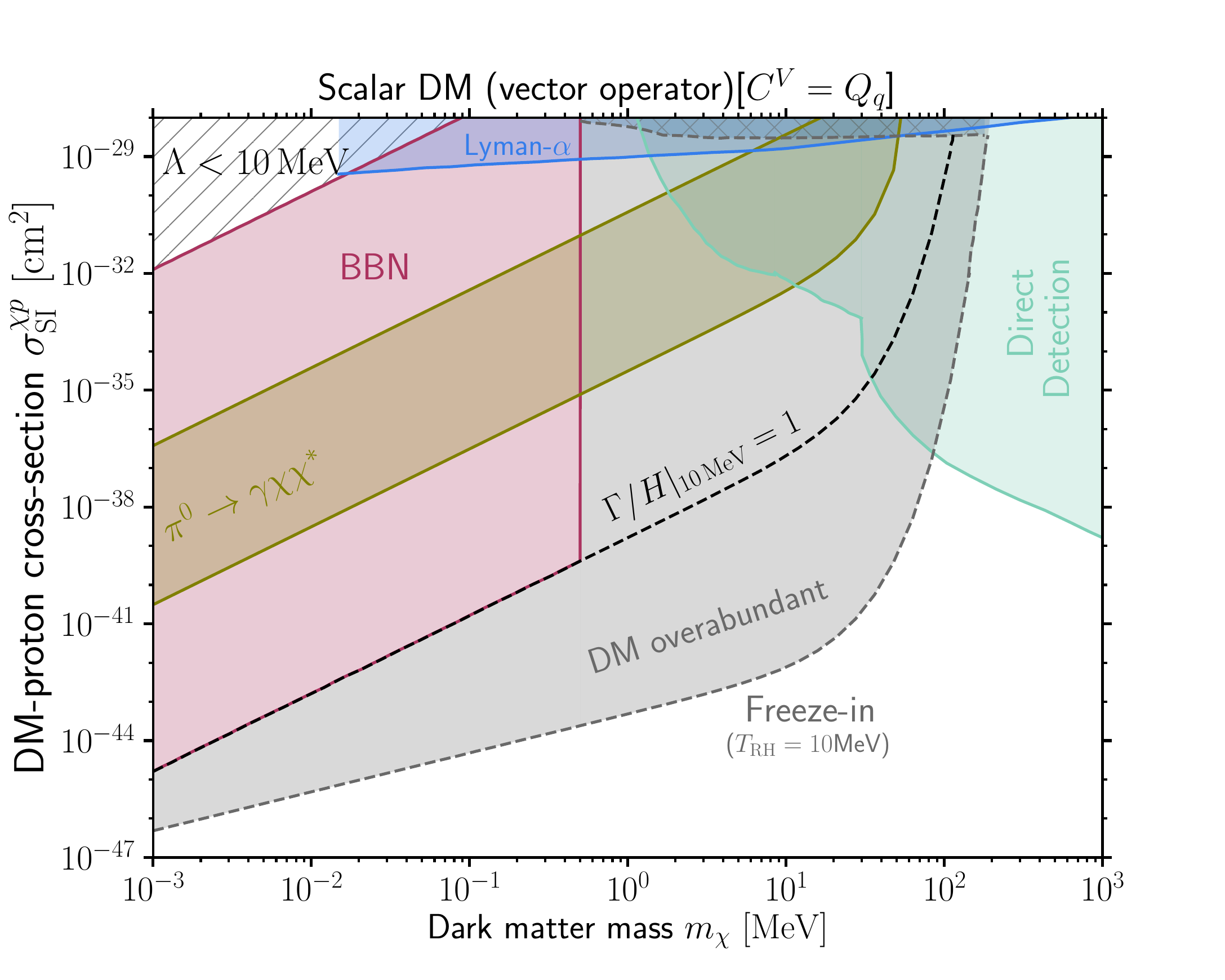}\!\!\!\!\!\!\!
    \includegraphics[width=0.50\linewidth]{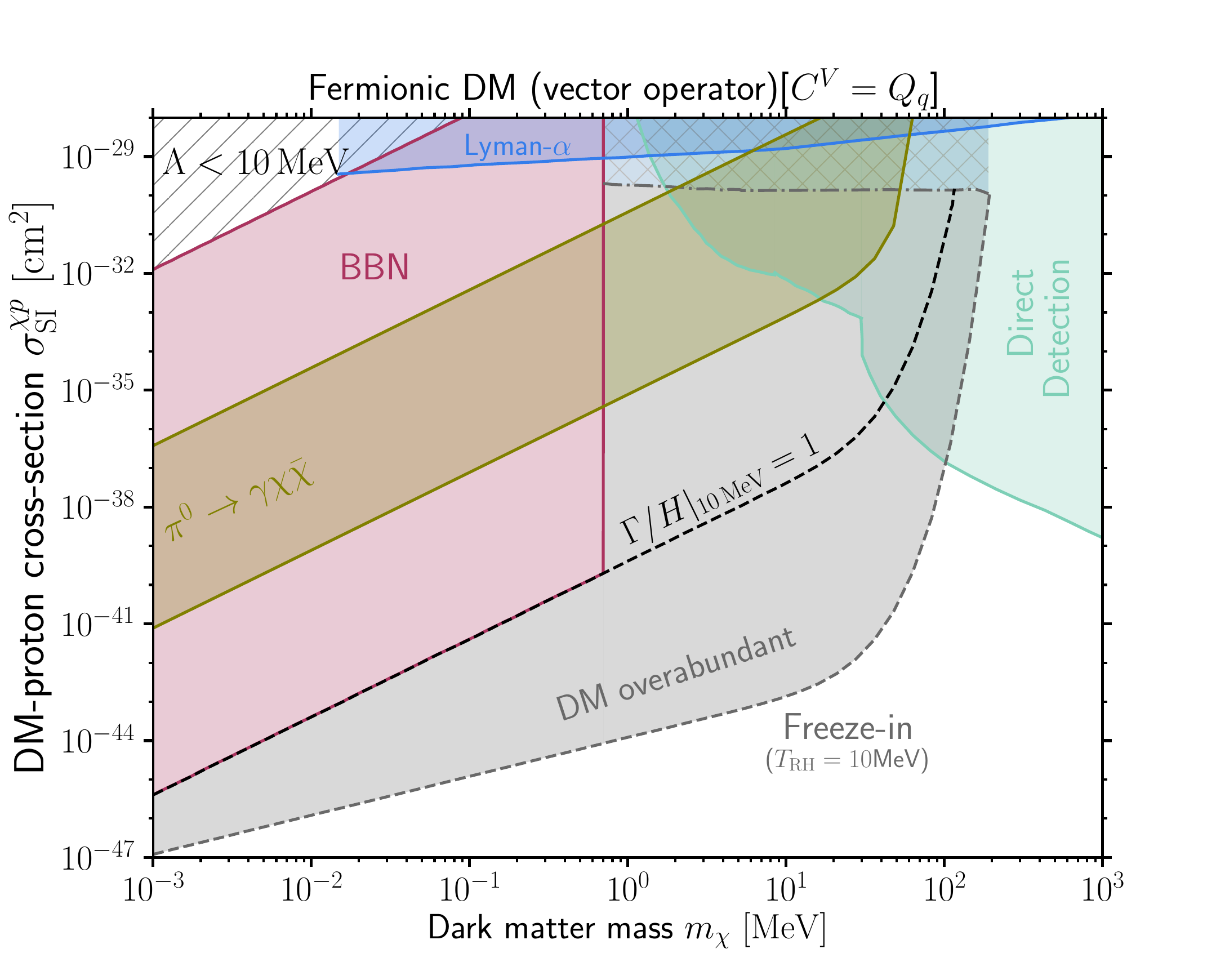}
    \caption{{\it Top:} Constraints on the DM-nucleon spin-independent cross-section $\sigma^{\chi N}_{\rm SI}$ for complex scalar (left-panel) and Dirac fermion (right-panel) DM that interacts with the quark vector current with $C^V = \mathbb{1}$. Red shaded regions are excluded by BBN and olive-green shaded regions by the upper-limit on the branching ratio of $\pi^0 \to \gamma \nu\bar{\nu}$. Above the black dashed lines the DM attains thermal equilibrium with the SM plasma below $T \sim 10\,{\rm MeV}$ and within the grey-shaded regions the irreducible DM abundance overcloses the Universe. {\it Bottom:} Same as top panel, but for $C^V = Q_q$. In this case, the DM freeze-out line is shown by the nearly flat dot-dashed grey line in the right panel, with the hatched region above excluded by CMB constraints~\cite{Slatyer:2015jla}. In all panels, the blue and light-green regions are excluded by existing bounds from structure formation~\cite{Rogers:2021byl} and direct-detection experiments~\cite{DAMIC-M:2025luv,SENSEI:2023zdf,PandaX:2023xgl}, respectively.}
    \label{fig:Vectorcouplingfigs}
\end{figure*}

\subsubsection{\texorpdfstring{$C^V = \mathbb{1}$}{CV=1}}

Let us first consider the case $C^V = \mathbb{1}$. This is the least constrained scenario, since the $\mathcal{O}(p^2)$ terms in the first two lines of the DM-meson interaction Lagrangian in \cref{eq:DMVcoupChPT} vanish. The $\mathcal{O}(p^4)$ term on the third line also vanishes and the DM thermalisation, freeze-in production, and relevant meson decay constraints therefore all arise from the WZW functional. 

As we have already mentioned, this interaction term leads to $\gamma\gamma \to \gamma \chi \bar{\chi}$ with a rate given in \cref{eq:Gggchichig} which simplifies to
\begin{align}
    \langle \Gamma_{\gamma\gamma \to \gamma \chi \bar{\chi}} \rangle \sim & \,\frac{\sigma_{\rm SI}^{\chi N}}{\mu^2_{\chi N}} \alpha^3 \frac{m^{10}_{\pi^0}}{\Gamma_{\pi^0} f^4} \left(\frac{m_{\pi^0}}{T}\right)^2 K_1\left(m_{\pi^0}/T\right) \,, 
\label{eq:wzwratesimple}
\end{align}
in the $T \gg m_\chi$ limit. Here, $K_1$ is a modified Bessel function of the second kind, which decays exponentially for large values of its argument; from \cref{eq:wzwratesimple}, it is clear this occurs for the temperatures of interest ($T < 10\,{\rm MeV}$). The DM thermalisation rate thus decreases exponentially with decreasing $T$; hence, if the DM is out-of-equilibrium at $T \sim 10\,{\rm MeV}$, it remains so at lower temperatures. We compare the interaction rate \eqref{eq:Gggchichig} against the Hubble rate at $T \sim 10\,{\rm MeV}$ to identify the region in the $\sigma^{\chi N}_{\rm SI}-m_\chi$ plane where the DM thermalises with the SM. This is delineated by the black dashed line in the top panels of \cref{fig:Vectorcouplingfigs} (scalar DM on the left and fermionic DM on the right). The red shaded regions, where the DM is in equilibrium at $T=10\,$MeV, are then excluded by BBN. Note that in the hatched region in the top-left corner of both panels, $\Lambda < 10\,$MeV and our EFT description is no longer valid at the relevant temperatures. In this region of parameter space, a light mediator ($m\lesssim10$\,MeV) is needed to achieve such large cross-sections, and BBN bounds on the mediator are expected to exclude this region.

The irreducible abundance of DM created via $\gamma\gamma \to \gamma \chi \bar{\chi}$ at temperatures below 10\,MeV saturates the observed DM abundance along the grey dashed curves in both panels; hence, in the regions between these curves and the black dashed lines, DM is overproduced via freeze-in. Above the black dashed lines, the DM abundance produced below 10\,MeV is dictated by thermal freeze-out (via $\chi \bar\chi \to \gamma \gamma\gamma$). The region where freeze-out saturates the observed DM abundance is above the top of the figures, and the grey shaded region denotes where the irreducible DM abundance (produced via either freeze-in or freeze-out) overcloses the Universe.

Since $\Gamma_{K^+ \to \pi^+ \chi \bar{\chi}}$ vanishes for $C^V = \mathbb{1}$ (see \cref{eq:Kaondecayvector}), the most relevant flavour physics constraint arises from the $\pi^0 \to \gamma \chi \bar{\chi}$ decay. The corresponding decay rate (given in \cref{eq:pi03bodydecays}) scales as 
\begin{equation}
    \Gamma_{\pi^0 \to \gamma \chi\bar{\chi}} \sim \,\frac{\sigma^{\chi N}_{\rm SI}}{\mu^2_{\chi N}} \alpha \frac{m^7_{\pi^0}}{f^2} \,,
    \label{eq:pi03bodydecaysmassless}
\end{equation}
in the limit $m_\chi \ll m_{\pi^0}$. The upper bound on the branching ratio of $\pi^0 \to \gamma\nu \bar{\nu}$~\cite{NA62:2019meo} excludes the olive-green shaded regions in \cref{fig:Vectorcouplingfigs}. The upper boundary of these regions corresponds to $\Lambda = m_{\pi^0}$, since beyond this line the EFT description becomes invalid. Of course, in a model that includes the light mediator, this region of parameter space is expected to be excluded, with the mediator produced on-shell in the $\pi^0$ decay.

In each panel, we also present the leading constraints from structure formation~\cite{Rogers:2021byl} in blue and direct-detection searches using the Migdal effect~\cite{DAMIC-M:2025luv,SENSEI:2023zdf,PandaX:2023xgl} in light-green. (As discussed in Ref.~\cite{Diamond:2023fsm}, there are also comparable direct detection constraints from loop-induced scattering with electrons.) Note that our constraints from BBN and, in particular, the irreducible DM abundance exceed the existing astrophysical and direct detection bounds by more than 10 orders of magnitude in $\sigma^{\chi N}_{\rm SI}$ for low DM masses.

\subsubsection{\texorpdfstring{$C^V = Q_q$}{CV=Qq}}

Next, we discuss the case $C^V = Q_q$, with our results shown in the lower panels of \cref{fig:Vectorcouplingfigs} (left panel for scalar DM and right for fermionic DM). Note that here the DM couples dominantly to protons (see \cref{eq:SIDDvectorcoup}) so that the constraints are now presented in the $m_\chi - \sigma^{\chi p}_{\rm SI}$ plane.

The main difference from the previous case is that here the leading order terms in the first two lines of the DM-meson Lagrangian \eqref{eq:DMVcoupChPT} are non-vanishing. The dominant DM thermalisation and production process is therefore $e^+e^- \to \chi \bar{\chi}$. Comparing $\langle \Gamma_{e^+ e^- \to \chi\bar{\chi}} \rangle$ (see \cref{eq:Geechichi}) with the Hubble rate at $T \sim 10\,{\rm MeV}$ identifies the regions of parameter space where the DM is in equilibrium at 10\,MeV (above the black dashed line) and which are excluded by BBN (red regions). In the limit $T \gg m_\chi,m_e$, the interaction rate $\langle \Gamma_{e^+ e^- \to \chi\bar{\chi}} \rangle $ scales as
\begin{equation}
    \langle \Gamma_{e^+ e^- \to \chi\bar{\chi}} \rangle \sim \, \frac{\sigma_{\rm SI}^{\chi p}}{\mu^2_{\chi N}} \alpha^2 T^5 \,,
    \label{eq:vecratesimple}
\end{equation}
where the $\mu^{-2}_{\chi N} \approx m_\chi^{-2}$ dependence explains why the BBN constraints become stronger at lower DM masses. 

The grey shaded regions again denote the parameter space where the irreducible DM abundance exceeds the observed abundance. Notice that this excludes significantly smaller cross-sections than in the previous $C^V=\mathbb{1}$ case, since here the production is via $e^+ e^- \to \chi\bar\chi$. Note that the freeze-in relic density scales as $m_\chi \langle \Gamma_{e^+ e^- \to \chi\bar\chi} \rangle  \approx m^{-1}_\chi$, explaining why the freeze-in contour has a shallower slope than the BBN constraint.

For fermionic DM, in the bottom-right panel of \cref{fig:Vectorcouplingfigs}, the nearly flat grey dot-dashed line denotes where the correct DM relic density is achieved via the freeze-out process $\bar{\chi}\chi \to e^+e^-$, assuming a maximum temperature of $10\,{\rm MeV}$.  While DM is under-produced above this line, such large $s$-wave annihilation cross-sections are excluded by the strong CMB constraints on energy injection during recombination~\cite{Slatyer:2015jla}. Note that for scalar DM (bottom-left panel of \cref{fig:Vectorcouplingfigs}), the corresponding region is also ruled out by the constraint from the Lyman-$\alpha$ forest.

The meson decay rate $\Gamma_{K^+ \to \pi^+ \chi \bar{\chi}}$ vanishes for $C^V = Q_q$ (see \cref{eq:Kaondecayvector}) and the strongest meson decay bound again comes from $\pi^0 \to \gamma\nu \bar{\nu}$, which excludes the olive-green regions.

In the four scenarios we have examined in \cref{fig:Vectorcouplingfigs}, scattering cross-sections larger than $10^{-36}\,\text{cm}^2$ are excluded, with significantly stronger constraints at low masses. The least constrained region that could be probed by direct detection experiments is for $m_\chi \gtrsim  50$~MeV. Finally, we re-emphasise that these overall conclusions apply more generally. In fact, for other choices of $C^V$, there will be constraints from $K^+$ decay that are comparable to the bound from DM overproduction.

\section{Axial-vector Operator}
\label{sec:axialvector}

In this section we consider DM interacting via the axial-vector quark current (third row of \cref{tab:operators}).  The leading order Chiral Lagrangian is given by the $C^A$-dependent term in \cref{eq:chiralEFT-quark}. Expanding this in terms of the meson fields, the leading interaction is with a single $\pi^0$ or $\eta$; however, at $\mathcal{O}(p^2)$ these couplings vanish for certain choices of $C^A$ and hence we also include the contributions from the $\mathcal{O}(p^4)$ Chiral Lagrangian~\cite{Gasser:1984gg}. The relevant Lagrangian is then
\begin{widetext}
\begin{align}
    \mathcal{L} &\supset -\frac{1}{\Lambda^2}(\bar\chi\gamma^\mu\gamma^5\chi)\,f \Bigg[ \Bigg(\left(C^A_u-C^A_d\right)\bigg[1+\frac{16 B_0 L_4}{f^2} \sum_{q=u,d,s}m_q\bigg] + \frac{16 B_0 L_5}{f^2}\left(C^A_u m_u - C^A_d m_d\right) \Bigg) \partial_\mu \pi^0 \nonumber\\
    & + \frac{1}{\sqrt{3}}\Bigg(\left(C^A_u+C^A_d-2C^A_s\right) \bigg[1+\frac{16 B_0 L_4}{f^2} \sum_{q=u,d,s}m_q\bigg] +   \frac{16 B_0 L_5}{f^2} \left(C^A_u m_u + C^A_d m_d - 2 C^A_s m_s\right) \Bigg)\partial_\mu \eta_8 \Bigg] \,,
    \label{eq:DMAVcoupChPT}
\end{align}
\end{widetext}
where $L_4 = 1.87 \times 10^{-3}$ and $L_5 = 1.22 \times 10^{-3}$ are low-energy constants~\cite{Bijnens:2014lea}. We neglect the $L_{4,5}$-dependent terms whenever the $\mathcal{O}(p^2)$ terms are non-vanishing. Note in particular that the $L_5$-dependent terms have a different flavour dependence in the $C^A_q$. For the axial-vector operator we only consider fermionic DM, since for scalar DM the interactions are CP-violating.  

The DM interactions with the quark axial-vector current lead to spin-dependent scattering between DM and nucleons (see e.g.~\cite{Cirelli:2013ufw}), with cross-section given by
\begin{equation}
    \sigma^{\chi N}_{\rm SD} = \frac{3\mu^2_{\chi N}}{\pi \Lambda^4} \bigg(\sum_{q=u,d,s} C^A_q \Delta^{(N)}_q \bigg)^2 \,,
    \label{eq:SDDDavectorcoup}
\end{equation}
where $\Delta^{(N)}_q$ are obtained from~\cite{Bishara:2017pfq}. To compare our results with direct-detection constraints we again trade the EFT scale $\Lambda$ for $\sigma^{\chi N}_{\rm SD}$ in the following discussion. Note that the differences in $\Delta^{(N)}_q$ for $N=p$ and $N=n$ are nominal and $ \sigma^{\chi N}_{\rm SD} \equiv \sigma^{\chi p}_{\rm SD} \simeq \sigma^{\chi n}_{\rm SD}$.

\begin{figure}[t]
    \centering
    \includegraphics[width=0.8\linewidth]{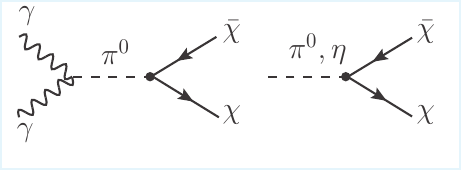}
    \caption{{\it Left:} Feynman diagram contributing to DM thermalisation and freeze-in production for interactions with the quark axial-vector, pseudoscalar and $G_{a\,\mu\nu}\tilde{G}^{a\,\mu\nu}$ operators. {\it Right:} Feynman diagram contributing to the invisible $\pi^0,\eta$ decays for interactions with the quark axial-vector, pseudoscalar and $G_{a\,\mu\nu}\tilde{G}^{a\,\mu\nu}$ operators.} 
    \label{fig:PSAVcoupFD}
\end{figure}

\begin{figure*}[t]
    \centering 
    \includegraphics[width=0.50\textwidth]{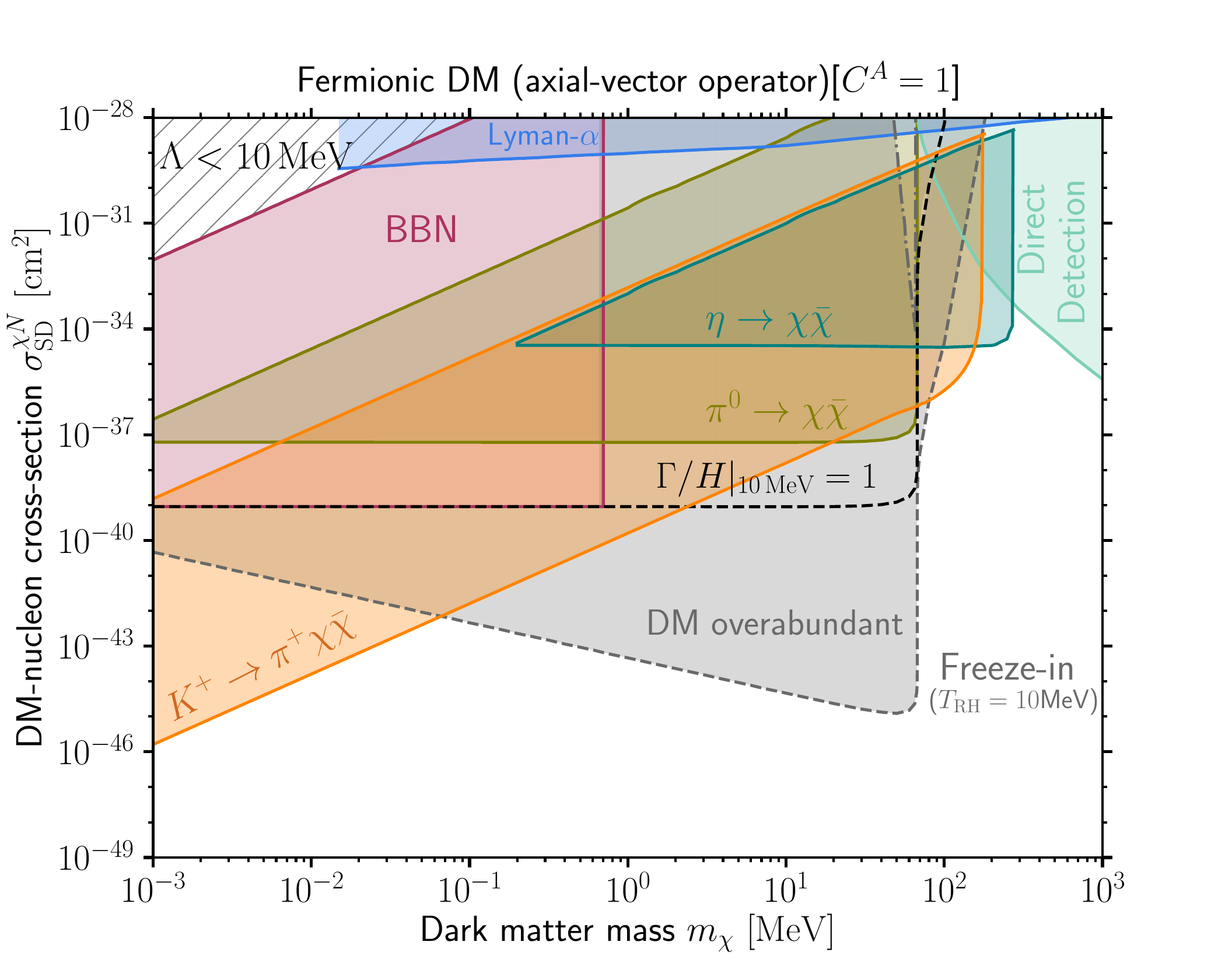}\!\!\!\!\!\!\!
    \includegraphics[width=0.50\textwidth]{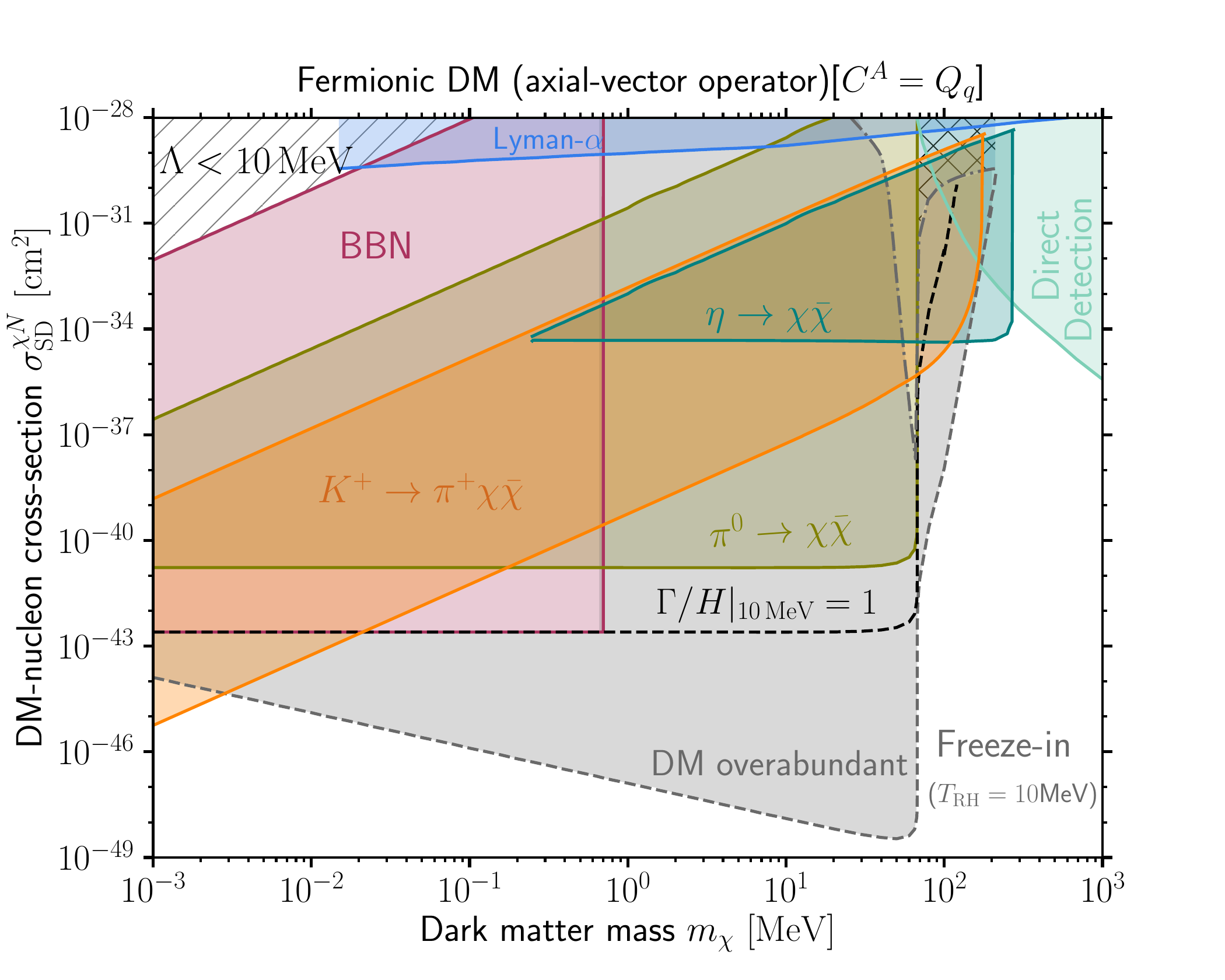}
    \caption{Constraints on the DM-nucleon spin-dependent cross-section $\sigma^{\chi N}_{\rm SD}$ for Dirac fermion DM interacting with the quark axial-vector current. The olive-green and deep-green shaded regions are excluded by the upper-limits on the $\pi^0 \to \nu\bar{\nu}$ and $\eta \to \nu\bar{\nu}$ branching ratios, respectively, and the orange shaded regions by the measurement of the $K^+ \to \pi^+ \nu\bar{\nu}$ branching ratio. The grey shaded regions denote the parameter space where the irreducible DM abundance overcloses the Universe, and above the black dashed lines the DM attains thermal equilibrium with the SM plasma below $T \sim 10\,{\rm MeV}$. Red shaded regions are excluded by BBN. The left panel is for $C^A = \mathbb{1}$ and the right panel $C^A = Q_q$. Existing bounds from structure formation~\cite{Rogers:2021byl} and direct-detection experiments~\cite{Wang:2021oha} are shown in blue and light-green, respectively.}
    \label{fig:AVectorcouplingfigs}
\end{figure*}

The dominant interaction between the DM and the SM plasma at low temperatures is $\gamma\gamma \to \chi\bar{\chi}$, which proceeds via the Feynman diagram shown in the left panel of \cref{fig:PSAVcoupFD}. We neglect the contribution of the $\eta$-mediated diagram since it is suppressed by a factor of $\mathcal{O}(m^4_\pi/m^4_\eta)$. The rate of this process (given in \cref{eq:AVcouprate}) simplifies to 
\begin{equation}
    \langle \Gamma_{\gamma\gamma \to \chi\bar{\chi}} \rangle \sim \sigma^{\chi N}_{\rm SD} \alpha^2 \frac{m^4_{\pi^0}}{\Gamma_{\pi^0}} \left(\frac{m_{\pi^0}}{T}\right)^2K_1\left( m_{\pi^0}/T\right) \,,
    \label{eq:axialvecratesimple}
\end{equation}
in the $T \gg m_\chi$ limit. This process determines whether the DM is in equilibrium below $T=10\,$MeV and is responsible for the constraints from BBN and DM overproduction.

The dominant meson decay constraint for the axial-vector operator arises from the $K^+ \to \pi^+ \chi\bar{\chi}$ decay. The relevant interactions come from the $\Delta S = 1$ Lagrangian in \cref{eq:DeltaS1Lagterm1} (see also \eqref{eq:DeltaS1transitionaxialvector}), with the decay rate (see \cref{eq:KaondecayAxialVector}) scaling as 
\begin{multline}
    \Gamma_{K^+ \to \pi^+ \chi \bar{\chi}} \sim \frac{\sigma^{\chi N}_{\rm SD}}{\mu^2_{\chi N}} m^5_{K^+} G^2_F |V_{ud}|^2 |V^{*\,}_{us}|^2 f^4 \\
    \times \bigg[\left(g_8 + 4g_{27}\right) C^A_u + \Big(g_8 + 2g^S_8 - \frac{4}{3} g_{27} \Big) \sum_q C^A_q\bigg]^2 \,,
\label{eq:kaondecayAVratesimple}
\end{multline}
for $m_{K^+} \gg m_{\pi^+},m_\chi$. While this decay rate may be suppressed for certain particular values of the $C^A_q$, such scenarios are clearly non-generic and likely challenging to UV complete. 

Additionally, there are two other relevant meson decay constraints: one from the upper-limit on the branching ratio of $\pi^0 \to \nu \bar{\nu}$~\cite{NA62:2020pwi} and the other from $\eta \to \nu \bar{\nu}$~\cite{NA64:2024mah}. The corresponding decays to DM proceed via the right-hand Feynman diagram in \cref{fig:PSAVcoupFD}. The decay rates (given in \cref{eq:pi0decayAV} and \cref{eq:etadecayAV}) parametrically scale as
\begin{equation}
    \Gamma_{i \to \chi \bar{\chi}} \sim \sigma^{\chi N}_{\rm SD} f^2 m_i \,,
    \label{eq:pietadecayAVsimple}
\end{equation}
in the limit $m_i \gg m_\chi$, with $i=\pi^0, \eta$.

\subsection{Models and Results}
\label{sec:axialvectormodels}

We present results for the same two flavour structures as considered previously in the vector case, $C^A=\mathbb{1}$ and $C^A=Q_q$. Once again, the flavour universal case is of particular interest since it corresponds to the least constrained scenario, with the $\mathcal{O}(p^2)$ terms in \cref{eq:DMAVcoupChPT} vanishing.

UV completions of the axial-vector interaction are far more involved than the simple UV completions we discussed in the vector case. The obvious starting point is a gauged $U(1)_X$ with axial charges for the SM quarks; however, for any such symmetry that is consistent with generating the quark masses from a single Higgs doublet, the new gauge boson has suppressed axial-vector couplings to the SM fermions (i.e. it couples to the quark vector current at low-energies)~\cite{Kahn:2016vjr}. UV completions therefore require a non-minimal Higgs sector or other quark mass generation mechanism. Another complication is that, for a flavour-universal symmetry, the fermions needed for anomaly cancellation must be coloured and so are subject to stringent collider bounds. These challenges are well known, and while it is possible to construct UV-complete models~\cite{Gori:2025jzu} (see also \cite{Kahn:2016vjr,Ismail:2016tod}), there are generally strong model-dependent constraints.

Putting aside the above caveats about UV completions, our results in the following sections show that there are in any case stringent low-energy bounds on the axial-vector interaction.

\subsubsection{\texorpdfstring{$C^A = \mathbb{1}$}{CA=1}}

We begin with the case of flavour-universal couplings, $C^A = \mathbb{1}$. This is the least constrained scenario, with all of the (non-vanishing) DM interactions in \cref{eq:DMAVcoupChPT} proportional to $L_5$ and suppressed by the quark masses. 

Our constraints on the spin-dependent DM-nucleon cross-section are shown in the left panel of \cref{fig:AVectorcouplingfigs}. The red shaded region is excluded by BBN, while in the grey shaded region the irreducible DM abundance produced via $\gamma \gamma \to \bar\chi \chi$ exceeds the observed abundance. Notice that the thermally-averaged rate $\langle \Gamma_{\gamma\gamma \to \chi\bar{\chi}} \rangle$ in \cref{eq:axialvecratesimple} does not depend on $m_\chi$. The BBN constraint is therefore independent of $m_\chi$, while the DM freeze-in relic density is $\propto m_\chi$ and the DM overabundance bound steadily strengthens with DM mass. 

The meson decays $K^+ \to \pi^+ \chi \bar\chi$, $\pi^0 \to \bar\chi \chi$, and $\eta \to \bar\chi \chi$ exclude the orange, olive-green, and deep-green regions, respectively. The upper edges of these regions occur where $\Lambda$ is equal to the mass of the decaying meson; above this line the mediator would be produced on-shell in the decay and the EFT can not be used to set constraints. Note that, when expressed in terms of $\sigma^{\chi N}_{\rm SD}$, the decay rate $\Gamma_{K^+ \to \pi^+ \chi \bar{\chi}}$ in \cref{eq:kaondecayAVratesimple} scales as $m^{-2}_\chi$ and the corresponding bound in \cref{fig:AVectorcouplingfigs} strengthens for lower DM masses. The upper-bounds on $\sigma^{\chi N}_{\rm SD}$ from $\pi^0$ and $\eta$ decays, on the other hand, are independent of the DM mass when $m_\chi \ll m_{\pi^0},m_\eta$ (see \cref{eq:pietadecayAVsimple}). 

The relevant direct detection constraints, obtained using a reinterpretation of XENON1T data~\cite{Wang:2021oha}, are shown in light-green and structure formation constraints~\cite{Rogers:2021byl} are shown in blue. Our constraints are more than 10 orders of magnitude stronger than these existing bounds for $m_\chi$ in the range 10 keV\,-\,100\,MeV.

\subsubsection{\texorpdfstring{$C^A = Q_q$}{CA=Qq}}

As our second representative choice, we consider $C^A = Q_q$, for which the terms $\propto L_{4,5}$ in \cref{eq:DMAVcoupChPT} are sub-leading and can be neglected. The form of the DM interactions, however, remains the same as in the $C^A=\mathbb{1}$ case; the only difference is that the couplings are enhanced by a factor of $\sim 10^{2}$. This leads to significantly stronger bounds from BBN, DM overproduction and $\pi^0$ decays. The $K^+ \to \pi^+ \bar\chi \chi$ and $\eta$ decay rates, on the other hand, are qualitatively similar to the $C^A=\mathbb{1}$ case, as can be seen from \cref{eq:kaondecayAVratesimple,eq:etadecayAV}. Our results are shown in the right panel of \cref{fig:AVectorcouplingfigs}.

The freeze-out relic density curve (grey dot-dashed) in the right panel of \cref{fig:AVectorcouplingfigs} shows a sharp dip around $m_\chi \sim 65\,{\rm MeV}$ which occurs because DM annihilation proceeds through the $\pi^0$ pole. The entire parameter region above this contour is ruled out by CMB, $\pi^0,\eta$ decays and direct detection.

\section{Pseudoscalar Operator}
\label{sec:pseudoscalar}

\begin{figure*}[t]
    \centering
    \includegraphics[width=0.50\linewidth]{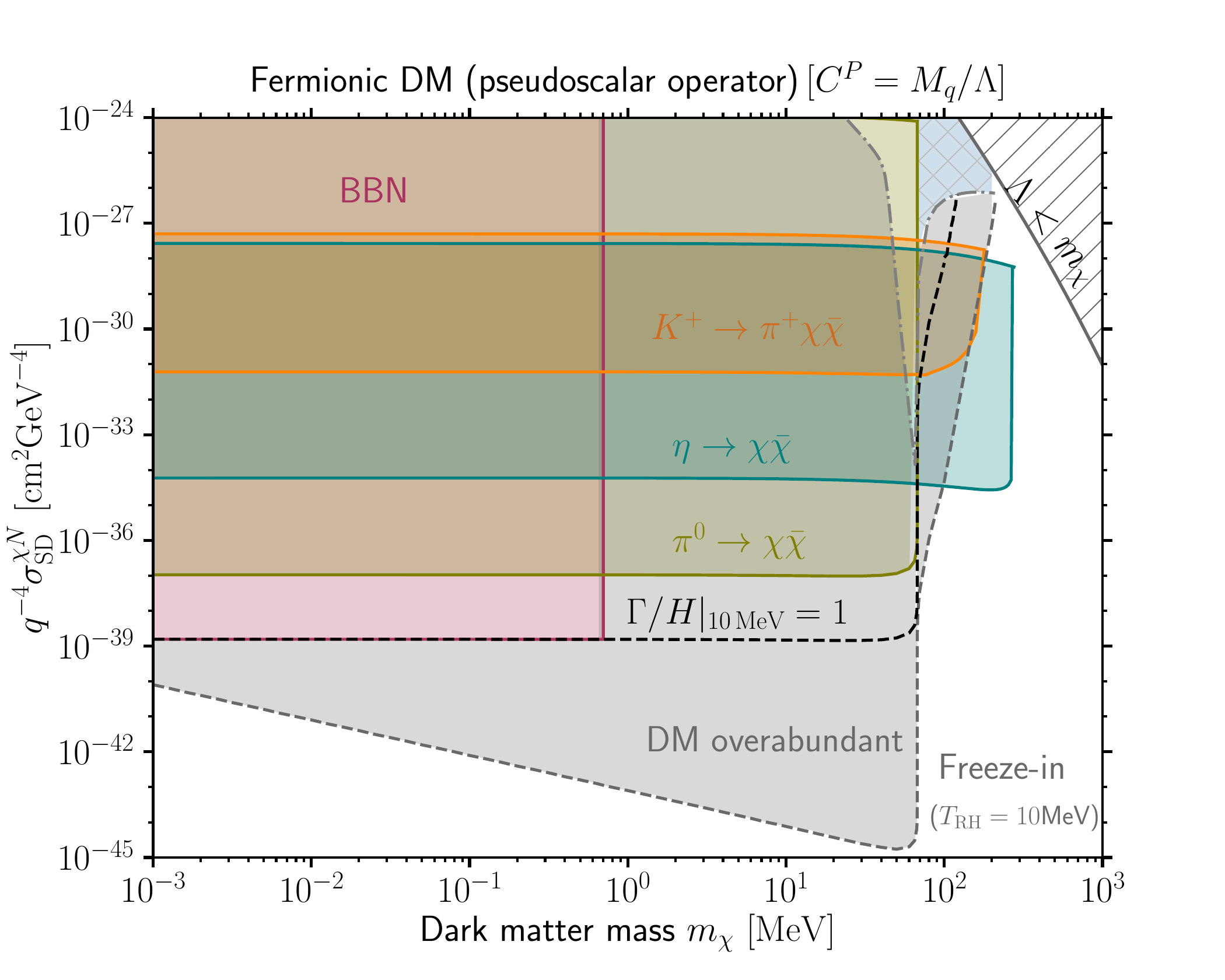}\!\!\!\!\!\!\!
    \includegraphics[width=0.50\linewidth]{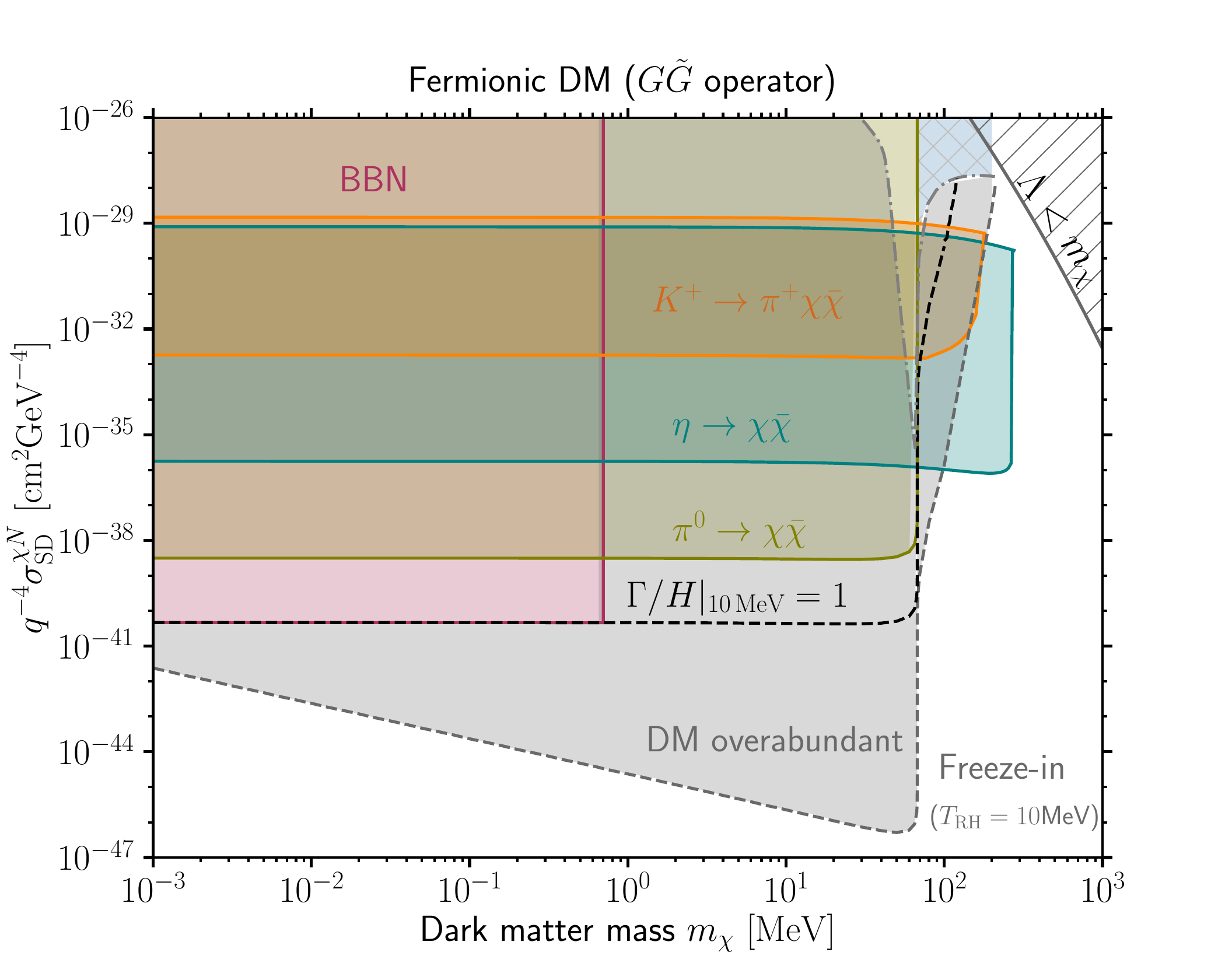}
    \caption{Constraints on the DM-nucleon spin-dependent cross-section $\sigma^{\chi N}_{\rm SD}$ (normalised by powers of the momentum-transfer $q$) for Dirac fermion DM interacting with the quark pseudoscalar operator (left panel) and $G_{a\mu\nu}\tilde{G}^{a\mu\nu}$ (right panel). The olive-green, deep-green and orange shaded regions are excluded by meson decays. The grey shaded regions denote where the irreducible DM abundance produced by $\gamma \gamma \to \chi\bar{\chi}$ overcloses the Universe. Above the black dashed lines DM attains thermal equilibrium with the SM plasma at $T \sim 10\,{\rm MeV}$, and the red region is excluded by BBN. The hatched grey region in the top-right corner of each panel denotes where our EFT description is no longer valid.}
    \label{fig:PSGGcouplingfigs}
\end{figure*}

In this section we present the constraints on DM that couples to the pseudoscalar quark operator (fourth row of \cref{tab:operators}). The leading order interactions are given by the $C^P$-dependent terms in \cref{eq:chiralEFT-quark}, which, expanded to linear order in the meson fields, give
\begin{multline}
    \mathcal{L} \supset \frac{1}{\Lambda^2} (\bar\chi i\gamma^5 \chi) B_0 f \\
    \times\bigg[(C^P_u-C^P_d)\pi^0 + \frac{(C^P_u+C^P_d-2 C^P_s)}{\sqrt{3}}\eta_8 \bigg]
    \label{eq:DMPScoupChPT}
\end{multline}
for fermionic DM. We do not consider scalar DM since the corresponding interaction violates CP.

For pseudoscalar interactions, the non-relativistic DM-nucleon scattering rate is spin-dependent and also suppressed by the momentum transfer $q \sim m_\chi v_\chi$, with $v_\chi\sim10^{-3}$ the non-relativistic velocity of the DM within the galactic halo. The cross-section is~\cite{Cirelli:2013ufw}
\begin{equation}
    \sigma^{\chi N}_{\rm SD} = \frac{g^2_N}{3\pi}\frac{q^4}{\Lambda^4}\frac{\mu^6_{\chi N}}{m^6_\chi} \,, 
    \label{eq:DDPSxsec}
\end{equation}
where
\begin{equation}
    g_N = \sum_{q=u,d,s} \frac{C^P_q}{m_q} \left( \Delta^{(N)}_q - \tilde{m} \sum_{q^\prime=u,d,s} \frac{\Delta^{(N)}_{q^\prime}}{m_{q^\prime}} \right) \,,
\end{equation}
with $\Delta^{(N)}_q$ taken from~\cite{Bishara:2017pfq} and $1/\tilde{m} = \sum_{q=u,d,s} 1/m_q$. The $q^4$-dependence of $\sigma^{\chi N}_{\rm SD}$ implies that neither direct-detection nor structure formation provide competitive constraints on this scenario. In this section, we express our results as constraints on the combined factor $q^{-4}\sigma^{\chi N}_{\rm SD}$.

Similar to the axial-vector interaction, the leading process for DM thermalisation and production at 10\,MeV temperatures is $\gamma\gamma \to \chi\bar{\chi}$, mediated by the neutral pion as shown in the left diagram of \cref{fig:PSAVcoupFD}. The DM thermalisation rate is obtained using the cross-section presented in \cref{eq:PScasexsec} and reduces to  
\begin{align}
    \langle \Gamma_{\gamma \gamma \to \chi \bar{\chi}} \rangle &\sim \, \frac{q^{-4} \sigma^{\chi N}_{\rm SD}}{g^2_N}  \alpha^2 B^2_0\frac{m^4_{\pi^0}}{\Gamma_{\pi^0}} \left(\frac{m_{\pi^0}}{T}\right)^2 K_{1}\left(\frac{m_{\pi^0}}{T}\right) \,,
\label{eq:PSratesimpler}
\end{align}
in the $T \gg m_\chi$ limit. 

The leading order DM-meson Lagrangian \eqref{eq:DMPScoupChPT} leads to $\pi^0 \to \chi\bar{\chi}$ and $\eta \to \chi\bar{\chi}$ decays. The decay rates are provided in \cref{eq:pi0decayPS,eq:etadecayPS}, respectively, and parametrically scale as 
\begin{equation}
    \Gamma_{i \to \chi \bar{\chi}} \sim \, \frac{q^{-4}\sigma^{\chi N}_{\rm SD}}{g^2_N} B^2_0 f^2
    m_i \,,
    \label{eq:pietadecayPSsimple}
\end{equation}
in the limit $m_i \gg m_\chi$. 

Unlike the vector and axial-vector cases, for the pseudoscalar operator, the $\Delta S = 1$ Lagrangian \eqref{eq:DeltaS1Lagterm1} does not contain additional interaction terms between the DM and the mesons. The decay $K^+ \to \pi^+ \chi \bar\chi$ therefore proceeds via the DM couplings to the $\pi^0$ and $\eta_8$ in \eqref{eq:DMPScoupChPT}, combined with the $\mathcal{O}(\Pi^3)$ term in the $\Delta S = 1$ Lagrangian (see diagram II of \cref{fig:PSAVcoupKaon} in \cref{app:flavour}). The resulting decay rate is given in \cref{eq:KaondecayPseudoscalarsimple}. Note that, following the cuts imposed in the NA62 analysis, we exclude the kinematic region in which the intermediate $\pi^0$ is on-shell when computing the bounds (see \cref{app:meson-AV} for details).

\subsection{Models and Results}
\label{sec:pseudoscalarmodels}

For the pseudoscalar operator, we present results for the case $C^P=M_q/\Lambda$.\footnote{Here, we assume couplings only to the light quarks. If the DM has pseudoscalar couplings to the heavy quarks in the UV, integrating them out will generate a contribution to the $C^{\tilde{G}}$ Wilson coefficient at low energies.} This choice is motivated by UV completions of this scenario and also respects minimal flavour violation. As was the case for the vector and axial-vector operators, the flavour-universal case $C^P=\mathbb{1}$ is strictly speaking the least constrained possibility (at the level of the EFT), due to the vanishing of the leading-order interactions in \cref{eq:DMPScoupChPT}. However, one-loop effects above the electroweak scale generically induce flavour off-diagonal couplings in the low-energy EFT~\cite{Dolan:2014ska}. The constraints obtained with exactly flavour-universal couplings ($C^P=\mathbb{1}$) are therefore unlikely to be representative of \emph{any} complete model and so we do not show them here.

Minimal UV completions of the pseudoscalar operator involve a new pseudoscalar mediator that couples to the DM. This mediator can couple to the SM quarks through mixing with the pseudoscalar Higgs in a two Higgs doublet model (see e.g.~\cite{Ipek:2014gua,Berlin:2015wwa}).

\subsubsection{\texorpdfstring{$C^P = M_q/\Lambda$}{CP=Mq/Lambda}}

For our choice of $C^P = M_q/\Lambda$, the constraints on the combination $q^{-4}\sigma^{\chi N}_{\rm SD}$ are shown in the left panel of \cref{fig:PSGGcouplingfigs}. Above the black dashed line, the process $\gamma\gamma \to \chi\bar{\chi}$ brings the DM into equilibrium with the SM at $T=10$\,MeV and the red region is then excluded by BBN. The grey dashed line denotes where the irreducible DM abundance produced via $\gamma\gamma \to \bar{\chi}\chi$ saturates the DM abundance, with the grey shaded region excluded by DM overabundance. This region is bounded from above by the contour for the correct freeze-out relic density (dot-dashed grey), beyond which strong CMB constraints~\cite{Slatyer:2015jla} apply (cyan hatched region). 

The constraints from the meson decays $\pi^0 \to \chi\bar{\chi}$, $\eta \to \chi\bar{\chi}$, and $K^+ \to \pi^+ \chi \bar\chi$ are shown by the olive-green, deep-green and orange regions, respectively. The upper edges of these regions correspond to where $\Lambda$ is equal to the relevant meson mass, beyond which constraints can not be calculated within the EFT. 

There are no bounds from either structure formation or direct detection on the parameter space shown in \cref{fig:PSGGcouplingfigs} due to the suppression of the DM-nucleon cross-section at non-relativistic velocities. Finally, the hatched region in the top-right corner denotes where $\Lambda < m_\chi$ and the EFT is no longer valid for any of the processes we consider.

\section{Gluon Operator}
\label{sec:gluon}

Lastly, we consider the interaction of DM with the CP-odd gluon operator $G_{a\mu\nu}\tilde{G}^{a\mu\nu}$ (see sixth row of \cref{tab:operators}). Note that this operator is redundant and in the UV it can always be removed via a quark field redefinition. This can be seen at the level of the Chiral Lagrangian in \cref{eq:chiralEFT-gluon} where (at $\mathcal{O}(\Lambda^{-3})$) the $C^{\tilde{G}}$-dependent terms could also be obtained via a particular choice of the axial-vector and pseudoscalar Wilson coefficients, $C^A$ and $C^P$. Nevertheless, since this operator arises naturally in UV completions with an axion-like-particle mediator or a pseudoscalar mediator that couples to the heavy quarks, we believe it is useful to also provide specific constraints on $C^{\tilde{G}}$.

Expanding the $C^{\tilde{G}}$-dependent terms in \cref{eq:chiralEFT-gluon} to linear order in the meson fields, the only non-zero contributions are from the term $\propto \Tr[M_q(U-U^\dagger)]$, giving
\begin{align}
    \mathcal{L} \supset & \,\frac{1}{3\Lambda^3}(\bar{\chi} i \gamma^5\chi) B_0 f \notag\\
    & \times \bigg[(m_u-m_d)\pi^0 + \frac{(m_u+m_d-2 m_s)}{\sqrt{3}}\eta_8 \bigg]
\label{eq:DMGGcoupChPT}
\end{align}
for fermionic DM. Here, we simply set $C^{\tilde{G}}=1$ without loss of generality (i.e. we absorb it into $\Lambda$), since it has no flavour structure. As mentioned earlier, we consider only the CP-conserving interactions and, hence, only fermionic DM. 

Comparing \cref{eq:DMGGcoupChPT} with the Lagrangian for the quark pseudoscalar operator in \cref{eq:DMPScoupChPT}, it is clear that they are equivalent with the replacement $C^P_q \to m_q/3\Lambda$ in \eqref{eq:DMPScoupChPT}. As a result, the constraints on DM interacting with $G_{a\mu\nu}\tilde{G}^{a\mu\nu}$ are the same as for the pseudoscalar operator with $C^P=M_q$, up to a trivial rescaling. The cross-section $\sigma_{\gamma \gamma \to \chi \bar\chi}$ and the decay rates for $K^+ \to \pi^+ \chi \bar\chi$, $\pi^0 \to \chi \bar\chi$, and $\eta \to \chi \bar\chi$ can therefore be obtained from \cref{eq:PScasexsec,eq:KaondecayPseudoscalar,eq:pi0decayPS,eq:etadecayPS} by taking $C^P_q \to m_q/3\Lambda$. 

The DM-nucleon scattering cross-section also takes the same form as for the quark pseudoscalar operator, with the replacement $g_N \to \tilde{g}_N$, where $\tilde{g}_N = -\tilde{m}\left( \sum_{q=u,d,s} \Delta^{(N)}_q/m_q \right)$~\cite{Cirelli:2013ufw}. The cross-section is therefore suppressed by a factor of $\tilde{g}^2_N/g^2_N \approx 3 \times 10^{-3}$ compared to that for the quark operator.

We show our results for the $G_{a\mu\nu}\tilde{G}^{a\mu\nu}$ operator in the right panel of \cref{fig:PSGGcouplingfigs}. As discussed above, these are a simply an overall rescaling of the constraints on the quark pseudoscalar operator with $C^P=M_q/\Lambda$ shown in the left panel. The bounds on the combination $q^{-4}\sigma^{\chi N}_{\rm SD}$ are a factor of $g^2_N/(9\tilde{g}^2_N) \sim 30$ stronger for the $G_{a\mu\nu}\tilde{G}^{a\mu\nu}$ operator.

\section{Summary and conclusion}
\label{sec:conclusion}

We have derived upper bounds on the direct detection cross-section for sub-GeV dark matter scattering off nucleons. The limits in this paper, when taken in conjunction with those of Ref.~\cite{Cox:2024rew} for scalar operators, indicate that direct detection experiments require sensitivity to cross-sections smaller than $10^{-36}\,\text{cm}^2$, if they are to probe unconstrained parameter space in this regime. This is several orders of magnitude lower than the often quoted astrophysical and cosmological bounds for these dark matter masses.

We have argued that our results are conservative and our conclusions apply independently of any specific UV model. We have derived bounds for scalar and vector operators (spin-independent scattering), and axial vector and pseudoscalar operators (spin-dependent scattering), in the low energy chiral effective theory. We have studied the most well-motivated coupling structures for these scenarios, arguing that other choices would be further constrained. 

Our bounds apply if the mediator mass is greater than a few hundred MeV. Lighter mediators are already known to lead to stringent constraints on the direct detection parameter space, due to bounds from meson decays, stars, and supernovae.

While we have studied dark matter with only hadronic couplings at leading order, we have shown that in all scenarios couplings to photons and electrons are generated at higher order. These couplings  produce an irreducible abundance of dark matter through freeze-in at low temperatures that can overclose the universe. Our bounds assume only that the universe reheated to a temperature of at least 10\,MeV, which is sufficient for viable BBN; higher maximum temperatures would lead to more dark matter production, and thus stronger bounds. 

Any UV model will be subject to additional constraints, beyond those we have considered here. These include strong bounds from $B$-meson decays and constraints from production of the mediator at the Large Hadron Collider.

\acknowledgments
This work was supported in part by the Australian Research Council through the ARC Centre of Excellence for Dark Matter Particle Physics, CE200100008. P.C. is supported by the Australian Research Council Discovery Early Career Researcher Award DE210100446.

\appendix
\onecolumngrid

\section{Chiral effective field theory}
\label{app:EFT}

In this appendix, we briefly review the derivation of the low-energy Chiral Lagrangian for the DM-meson system, starting from the interactions with quarks and gluons. The DM operators are included as external sources in the QCD Lagrangian which transform under the $SU(3)_L \times SU(3)_R$ chiral symmetry. This is then matched onto the low-energy Chiral Lagrangian~\cite{Gasser:1984gg,Bishara:2016hek}. 

We begin with the QCD Lagrangian in the presence of external sources,
\begin{equation} \label{eq:QCD}
    \mathcal{L} = \mathcal{L}^0_\text{QCD} + \bar{q} \gamma^\mu \left( v_\mu(x) + \gamma^5 a_\mu(x) \right) q - \bar{q} \left( s(x) - i \gamma^5 p(x) \right) q + s_G(x) \frac{\alpha_s}{8\pi} G^{a,\mu\nu}G_{a,\mu\nu} \,,
\end{equation}
where $\mathcal{L}^0_\text{QCD}$ is the QCD Lagrangian in the limit of zero quark masses and $s(x),p(x),v_\mu(x),a_\mu(x)$, and $s_G(x)$ are sources for the respective quark and gluon operators. We have omitted a source term for the $\tilde{G}^{a,\mu\nu}G_{a,\mu\nu}$ operator since it can be removed via a quark field redefinition and absorbed into $a_\mu(x)$ and $p(x)$.

The sources transform as spurions under the $SU(3)_L \times SU(3)_R$ symmetry according to
\begin{align} \label{eq:spurions}
    s + ip &\to V_R \left(s + ip \right) V_L^\dagger \,, \notag \\
    v_\mu + a_\mu &\to V_R \left( v_\mu + a_\mu \right) V_R^\dagger + i V_R \partial_\mu V_R^\dagger \,, \notag \\
    v_\mu - a_\mu &\to V_L \left( v_\mu - a_\mu \right) V_L^\dagger + i V_L \partial_\mu V_L^\dagger \,,\notag\\
    s_G &\to s_G \,,
\end{align}
where $V_{L,R}$ are the $3\times3$ transformation matrices. Given these transformation properties, the lowest order, i.e. $\mathcal{O}(p^2)$, Chiral Lagrangian in the presence of the external sources is given by
\begin{multline}\label{eq:ChPTOriginal}
    \mathcal{L}_{\mathcal{O}(p^2)} = \frac{f^2}{4} \Tr[\mathcal{D}^\mu U^\dagger \mathcal{D}_\mu U] + \frac{1}{2} B_0 f^2 \Tr[ \left( s(x) + ip(x) \right) U^\dagger + \text{h.c.} ] + s_G(x)  \frac{f^2}{18} \Big(\Tr[\partial^\mu U^\dagger \partial_\mu U] + 3 B_0 \Tr[s(x) (U+U^\dagger)] \Big) \,,
\end{multline}
where $U=\exp{(i\sqrt{2}\Pi/f)}$, and 
\begin{equation}
    \Pi = 
    \begin{pmatrix}
        \frac{\pi^0}{\sqrt{2}} + \frac{\eta_8}{\sqrt{6}} & \pi^+ & K^+ \\
        \pi^- & -\frac{\pi^0}{\sqrt{2}} + \frac{\eta_8}{\sqrt{6}} & K^0 \\
        K^- & \bar{K}^0 & \frac{-2\eta_8}{\sqrt{6}}
    \end{pmatrix}
\end{equation}
is the hermitian matrix of meson fields, with $f\approx92\,$MeV the pion decay constant at leading order, and $B_0=m_\pi^2/(m_u+m_d)$. Note that we do not include the $\eta^\prime$ and $\det U=1$. The covariant derivative acts as
\begin{equation}\label{eq:covDeriv}
    \mathcal{D}_\mu U = \partial_\mu U - i\left( v_\mu + a^\mu \right) U + iU \left( v_\mu - a_\mu \right) \,.
\end{equation}

Using \cref{eq:covDeriv}, we expand the Lagrangian~\eqref{eq:ChPTOriginal} and obtain
\begin{align}\label{eq:psqChPTMain}
    \mathcal{L}_{\mathcal{O}(p^2)} = 
    &-\frac{if^2}{2} \Big( \Tr[ v^\mu ( U \partial_\mu U^\dagger + U^\dagger \partial_\mu U)] - 2i \Tr[ v^\mu U v_\mu U^\dagger] + \Tr[a^\mu ( U \partial_\mu U^\dagger - U^\dagger \partial_\mu U) ] \Big)\notag \\
    &- \frac{iB_0f^2}{2} \Tr[p (U-U^\dagger)] + \frac{B_0 f^2}{2} \Tr[s (U+U^\dagger) ] + s_G  \frac{f^2}{18} \left(\Tr[\partial^\mu U^\dagger \partial_\mu U] + 3 B_0 \Tr[s (U+U^\dagger)] \right) \,,
\end{align}
The sources contain the DM field bilinears and are given by
\begin{align}\label{eq:DMbilinears}
    s(x) &= M_q + \frac{C^S}{\Lambda^2}
    \begin{cases}
        \bar\chi \chi  \quad \text{(fermionic DM)} \,, \\
        \chi^* \chi \quad \text{(scalar DM)} \,,
    \end{cases} \\
    p(x) &= \bigg( \frac{C^P}{\Lambda^2} + \frac{C^{\tilde{G}}}{3\Lambda^3} M_q \bigg) i \bar\chi \gamma^5 \chi \,, \\
    v^\mu(x) &= e Q_q A_\mu + \frac{C^V}{\Lambda^2}
    \begin{cases}
         \bar\chi \gamma^\mu \chi \quad \text{(fermionic DM)} \,, \\
         \chi^* \overset{\leftrightarrow}{\partial^\mu}\chi \quad \text{(scalar DM)} \,,
    \end{cases} \\
    a^\mu(x) &= \frac{C^A}{\Lambda^2} \bar\chi \gamma^\mu \gamma^5 \chi + \frac{C^{\tilde{G}}}{6\Lambda^2} \partial^\mu(i \bar\chi \gamma^5 \chi) \,.\\
    s_G(x) &= \frac{C^G}{\Lambda^3}  
    \begin{cases}
         \bar\chi \chi  \quad \text{(fermionic DM)} \,, \\
         \chi^* \chi \quad \text{(scalar DM)} \,.
    \end{cases}
\end{align}
Note that $s(x)$ also includes the quark masses and $v^\mu(x)$ the electromagnetic interactions. Substituting these expressions for the sources into \cref{eq:psqChPTMain} yields the DM-meson Lagrangians in \cref{eq:chiralEFT-quark,eq:chiralEFT-gluon}. 

For certain observables (e.g. $e^+e^- \to \chi\bar\chi$) or for flavour-universal Wilson coefficients, it becomes necessary to work at $\mathcal{O}(p^4)$ in the chiral counting. We refer the reader to Refs.~\cite{Gasser:1984gg,Pich:1995bw} for a detailed discussion of the $\mathcal{O}(p^4)$ Chiral Lagrangian. Finally, for $K^+ \to \pi^+ \chi \bar\chi$ decays, the $s \to d$ transition is governed by the $\Delta S = 1$ Lagrangian, as discussed in \cref{sec:mesondecay}.

\section{DM interaction rates and irreducible Freeze-in abundance}
\label{app:thermalisation}

In this appendix, we provide the relevant formulae for calculating the DM interaction rates in the early universe and the irreducible freeze-in abundance.  

\subsection{Vector operator} 

For the vector operator, DM interacts with the SM plasma at $T \lesssim 10\,{\rm MeV}$ via the processes $e^+ e^- \to \chi \bar{\chi}$ and/or $\gamma \gamma \to \gamma \chi \bar{\chi}$. The former process generally dominates, since the latter is suppressed by an additional factor of $\alpha$ and the three-body phase space. However, $\gamma \gamma \to \gamma \chi \bar{\chi}$ becomes the dominant process for the case of flavour-universal couplings, $C^V=\mathbb{1}$. In the following subsections we provide the thermally-averaged interaction rates for these processes and the corresponding freeze-in abundances.   

\subsubsection{\texorpdfstring{$e^+ e^- \to \chi \bar{\chi}$}{..}}

The process $e^+ e^- \to \chi \bar{\chi}$ proceeds via the diagrams shown in \cref{fig:EEChiChi}, with the tree-level diagram being proportional to $L_{10}+2H_1$. Combining these diagrams, we obtain the amplitude for $e^+ e^- \to \chi \bar{\chi}$ (assuming fermionic DM):
\begin{equation}
    i\mathcal{M}_{e^+e^- \to \chi \bar{\chi}} = \frac{i\,\alpha}{4\,\pi\,\Lambda^2} \left(\sum_{i=\pi^+,K^+}\!\!\!\mathcal{C}_i \big[ F_i(s) - \frac{64\pi^2}{3}(L_{10}+2H_1) \big] \right)  \bar v_e(k_2) \gamma^\nu u_e(k_1)  D_{\mu\nu}(\omega,s)  \left[ P^\alpha P^\mu - P^2\,\eta^{\alpha\mu}\right]  \bar u_\chi(p_1) \gamma_\alpha v_\chi(p_2) \,,
    \label{eq:Vcoupamp}
\end{equation}
where $P^\mu=k^\mu_1+k^\mu_2$ and, for convenience, we have defined $\mathcal{C}_\pi = C^V_u-C^V_d$ and $\mathcal{C}_K = C^V_u-C^V_s$. Here, the loop function is
\begin{equation}
    F_i(s) =\frac{8}{9} \left(1-\frac{3m^2_i}{s}\right) + \frac{1}{3} \log \left(\frac{\mu^2}{m^2_i} \right) + \frac{1}{3} \left(1-\frac{4m^2_i}{s} \right)^{3/2}\log \left[\frac{2m^2_i-s+\sqrt{s(s-4m^2_i)}}{2m^2_i} \right] \,,
    \label{eq:Vcouploop}
\end{equation}
with $i=\pi^+,K^+$ and the renormalisation scale $\mu = 10\,{\rm MeV}$. The UV-finiteness of $F_i(s)$ is ensured by including the counterterms for $L_{10}$ and $H_1$. Note that the corresponding amplitude for scalar DM is obtained by replacing $\bar u_\chi(p_1) \gamma_\alpha v_\chi(p_2) \to (p_1-p_2)_\alpha$ in \cref{eq:Vcoupamp}.

In our calculation, we have used the thermal photon propagator (see, e.g.~\cite{Scherer:2024uui})
\begin{align}
    D_{\mu\nu}(\omega,s) = & D_L P_{L,\mu\,\nu} + D_T P_{T,\mu\,\nu}, \qquad D_{L,T} = \frac{i}{s-\Pi_{L,T}} \,,
\end{align}
where $P_{L,\mu\,\nu}$ and $P_{T,\mu\,\nu}$ are longitudinal and transverse projection operators and $\Pi_{L,T}$ denote the photon polarisation functions at finite temperature. We have used the expressions for ${\rm Re}\,\Pi_{L,T}$ and ${\rm Im}\,\Pi_{L,T}$ given in Ref.~\cite{Scherer:2024uui} for our calculations. The importance of taking the photon thermal propagator into account lies in the fact that the photon acquires a plasma mass and hence a longitudinal polarisation. The effect of this longitudinal mode on DM cosmology is known to be non-negligible for sub-MeV DM masses~\cite{Dvorkin:2019zdi}.    

Using the amplitude $\mathcal{M}_{e^+e^- \to \chi \bar{\chi}}$ in \cref{eq:Vcoupamp}, and averaging (summing) over initial (final) state spins, we obtain the cross-section for $e^+ e^- \to \chi \bar{\chi}$,
\begin{align}
    \sigma_{e^+e^- \to \chi \bar{\chi}}(s,\omega) =  \frac{\alpha^2}{768\pi^3 \Lambda^4} \, \Bigg| \left(\sum_{i=\pi^+,K^+}\!\!\!\mathcal{C}_i \bigg[ F_i(s) - \frac{64\pi^2}{3}(L_{10}+2H_1) \bigg] \right) \Bigg|^2 \left[|D_T(\omega,s)|^2+\frac{2m^2_e}{s}|D_L(\omega,s)|^2 \right] & \, \, \,\notag\\
    \times \,\,\, s^3\sqrt{\frac{s-4m^2_\chi}{s-4m^2_e}}   
    \begin{cases}
     4\left( 1 + \frac{2m^2_\chi}{s}\right)\quad \text{(fermionic DM)} \,, \\
     \left( 1 - \frac{4m^2_\chi}{s}\right) \quad \text{(scalar DM)} \,,
    \end{cases} &
    \label{eq:vectorxsec}
\end{align}
and the corresponding interaction rate,
\begin{equation}
    \langle \Gamma_{e^+ e^- \to \chi\bar{\chi}} \rangle = \frac{1}{32\pi^4\,n_e(T)}\int_{\text{Max}[4m^2_\chi,4m^2_e]}^\infty ds (s-4m^2_e) \int_{\sqrt{s}}^\infty d\omega e^{-\omega/T}\sqrt{\omega^2-s}\,\sigma_{e^+e^- \to \chi \bar{\chi}}(s,\omega) \,,
    \label{eq:Geechichi}
\end{equation}
where $m_e$ is the electron mass and $n_e(T) = m^2_e T K_2(m_e/T)/\pi^2$ is the electron thermal number density. We find that at $T = 10\,{\rm MeV}$ the interaction rate $\langle \Gamma_{e^+ e^- \to \chi\bar{\chi}} \rangle$ obtained using the thermal photon propagator differs by less than a few percent from the rate obtained using the zero-temperature photon propagator. This is because the dominant contribution to $\langle \Gamma_{e^+ e^- \to \chi\bar{\chi}} \rangle$ comes from the region of phase space that does not contain the poles of the thermal photon propagator.

The DM yield, $Y_\chi$, produced via freeze-in is obtained by solving the Boltzmann equation
\begin{equation}
    \frac{dY_\chi}{dT} = - \frac{n_e(T)\langle \Gamma_{e^+ e^- \to \chi\bar{\chi}} \rangle}{s\,H\,T} \,,
    \label{eq:Botzmanneq}
\end{equation}
with the interaction rate $\langle \Gamma_{e^+ e^- \to \chi\bar{\chi}} \rangle$ given in \cref{eq:Geechichi} and where $s$ is the entropy density at temperature $T$. The irreducible DM freeze-in abundance produced from $e^+ e^- \to \chi\bar{\chi}$ is then
\begin{equation}
    \Omega_\chi h^2 = \frac{2025}{2\pi^4}\sqrt{\frac{2\pi^2}{45}} \frac{s_0}{\rho_c} \int_{T_0}^{T=10\,{\rm MeV}} dT \frac{m_\chi M_{\rm Pl}}{T^6} \frac{n_e(T)}{g_{*,S}(T)\sqrt{g_{*,\rho}(T)}} \langle \Gamma_{e^+ e^- \to \chi\bar{\chi}} \rangle \,,
    \label{eq:FIvectorcoup}
\end{equation}
where $s_0$ is the entropy density today, $\rho_c$ is the critical density of the Universe today, and $g_{*,S}(T)$ and $g_{*,\rho}(T)$ represent the number of relativistic d.o.f. contributing to the entropy density and energy density of the Universe at temperature $T$. The upper-limit of the integration follows from our assumption that the Universe reached a temperature of at least 10 MeV, while $T_0$ is the temperature of the present-day Universe.

\subsubsection{\texorpdfstring{$\gamma\gamma \to \gamma\chi \bar{\chi} $}{..}}

For flavour universal couplings, $\gamma\gamma \to \gamma\chi \bar{\chi}$ becomes the leading process through which the DM interacts with the SM plasma at $T \lesssim 10\,{\rm MeV}$, since only the WZW term in \cref{eq:DMVcoupChPT} is non-zero. This proceeds via the Feynman diagram shown in the left panel of \cref{fig:VcoupMeson}. There exists $t$ and $u$-channel diagrams as well, but these are sub-dominant at the temperatures of interest and, hence, we neglect them. Considering only the $s$-channel diagram, we obtain the amplitude (for fermionic DM)
\begin{equation} 
    i\mathcal{M}_{\gamma\gamma \to \gamma \chi \bar{\chi} } = \frac{i \alpha^{3/2}_{\rm EM}}{2 \pi^{3/2} f^2 \Lambda^2} \left( 2C^V_u+C^V_d\right)\epsilon^{\mu\nu\rho\sigma} \epsilon^{a b c d}  \partial_\rho (\bar u_\chi(p_1) \gamma_\sigma v_\chi(p_2)) \frac{\varepsilon_c(k_1)  \varepsilon_d(k_2) \varepsilon^*_\nu(k_3)k_{1 a} k_{2 b} k_{3\mu}}{s-m^2_{\pi^0}+i m_{\pi^0}\Gamma_{\pi^0}} \,,
    \label{eq:VcoupampWZW}
\end{equation}
where $\varepsilon$'s are photon polarisation vectors, $m_{\pi^0}$ is the $\pi^0$ mass and $\Gamma_{\pi^0}$ is its decay width. Note that here we simply use the zero-temperature propagator for the pion, since for $T \lesssim 10\,{\rm MeV}$ the finite-temperature corrections to the pion mass and decay width are negligible (see e.g.~\cite{Song:1994de}). 

Using the above expression for $\mathcal{M}_{\gamma\gamma \to \gamma \chi \bar{\chi}}$, we obtain the following cross-section for $\gamma\gamma \to \gamma \chi \bar{\chi}$,
\begin{equation}
    \sigma_{\gamma\gamma \to \gamma \chi \bar{\chi}}(s) = \frac{\alpha^3 \left( 2C^V_u+C^V_d\right)^2}{8192 \pi^8 f^4 \Lambda^4} \frac{s^5}{(s-m^2_{\pi^0})^2+m^2_{\pi^0}\Gamma^2_{\pi^0}} \mathcal{F}^V_{\gamma \gamma \to \gamma \chi \bar{\chi}} \,, 
\label{eq:vectorxsecWZW}
\end{equation}
where
\begin{align}
    \mathcal{F}^V_{\gamma \gamma \to \gamma \chi \bar{\chi}} = & \int_0^{1-\frac{4m^2_\chi}{s}}dx  \int_{y_{-}(x)}^{y_{+}(x)} dy \times 
    \begin{cases}
        \frac{2m^2_\chi}{s}x^2 + 2 (1-x) (1+y^2) - (1-x)(2-x)(x+2y) \quad \text{(fermionic DM)} \,, \\
        -\left[\frac{m^2_\chi}{s}x^2 + (1-x) (1-y) (1-x-y)\right] \quad \text{(scalar DM)} \,, \\
    \end{cases}
\end{align}
with $x = 2E_\gamma/\sqrt{s}$ the energy fraction carried by the final state photon, $y = 2E_\chi/\sqrt{s}$ the energy fraction carried by each of the $\chi$ and $\bar\chi$, and $y_{\pm}(x) = \frac{1}{2}\bigg[ (2-x) \pm x\sqrt{1-\frac{4m^2_\chi/s}{1-x}}\bigg]$. The resulting thermally-averaged interaction rate is
\begin{equation}
    \langle \Gamma_{\gamma\gamma \to \gamma \chi \bar{\chi}}\rangle = \frac{T}{16\pi^4 n_\gamma(T)} \int_{4m^2_\chi}^\infty ds \, s^{3/2} \sigma_{\gamma\gamma \to \gamma \chi \bar{\chi}}(s) K_1\big(\sqrt{s}/T\big) \,,
    \label{eq:Gggchichig}
\end{equation}
where $n_\gamma(T) = 2T^3/\pi^2$ is the photon thermal number density and $K_1(\sqrt{s}/T)$ is the modified Bessel function of the second kind. Solving the freeze-in Boltzmann equation, we obtain the irreducible DM freeze-in abundance
\begin{equation}
    \Omega_\chi h^2 = \frac{2025}{2\pi^4}\sqrt{\frac{2\pi^2}{45}} \frac{s_0}{\rho_c}  \int_{T_0}^{T = 10\,{\rm MeV}} dT \, \frac{m_\chi M_{\rm Pl}}{T^6} \frac{n_\gamma(T)}{g_{*,S}(T)\sqrt{g_{*,\rho}(T)}} \langle \Gamma_{\gamma\gamma \to \gamma \chi \bar{\chi}} \rangle \,.
    \label{eq:FIvectorcoupWZW}
\end{equation}

\subsection{Axial-vector operator} 

For the axial-vector operator, the DM interacts with the SM plasma via $\gamma \gamma \to \chi \bar{\chi}$, which proceeds through the Feynman diagram shown in the left panel of \cref{fig:PSAVcoupFD}. The amplitude for this process is readily obtained from the DM-meson Lagrangian in \cref{eq:DMAVcoupChPT} and is given as follows:
\begin{align}
    i\mathcal{M}_{\gamma\gamma \to \chi\bar{\chi}} &= \frac{i\alpha}{\pi \Lambda^2}\, \frac{\epsilon^{\mu\nu\alpha\beta}\,k_{1\mu}k_{2\nu}\epsilon_\alpha(k_1)\epsilon_\beta(k_2)}{s-m^2_{\pi^0}+i m_{\pi^0}\Gamma_{\pi^0}} P_\rho \bar u_\chi(p_1) \gamma^\rho \gamma^5 v_\chi(p_2) \notag\\
    & \times
    \left[ (C^A_u-C^A_d)\bigg(1+\frac{16 B_0 L_4}{f^2} \sum_q m_q \bigg) + \frac{16 B_0 L_5}{f^2} (C^A_u m_u- C^A_d m_d)\right] \,,
    \label{eq:AVcoupamp}
\end{align}
where $P_\rho = k_{1\rho}+k_{2\rho}$. The corresponding cross-section is
\begin{equation}
    \sigma_{\gamma\gamma \to \chi\bar{\chi}}(s) = \frac{\alpha^2\,m^2_\chi}{16\pi^3 \Lambda^4} \sqrt{1-\frac{4m^2_\chi}{s}} \frac{s^2}{(s-m^2_{\pi^0})^2+m^2_{\pi^0}\Gamma^2_{\pi^0}}\left[ (C^A_u-C^A_d)\bigg(1+\frac{16 B_0 L_4}{f^2} \sum_q m_q \bigg) + \frac{16 B_0 L_5}{f^2} (C^A_u m_u- C^A_d m_d)\right]^2 \,. 
    \label{eq:Axialvectorxsec}
\end{equation}
Using this cross-section we obtain the thermally-averaged interaction rate
\begin{equation}
    \langle \Gamma_{\gamma\gamma \to \chi\bar{\chi}} \rangle = \frac{T}{16\pi^4\,n_\gamma(T)} \int_{4m^2_\chi}^\infty ds \, s^{3/2}  \sigma_{\gamma\gamma \to \chi\bar{\chi}}(s) K_1\left(\sqrt{s}/T\right) \,,
    \label{eq:AVcouprate}
\end{equation}
and consequently the DM freeze-in abundance as
\begin{equation}
    \Omega_\chi h^2 = \frac{2025}{2\pi^4}\sqrt{\frac{2\pi^2}{45}} \frac{s_0}{\rho_c} \int_{T_0}^{T=10\,{\rm MeV}} dT \, \frac{m_\chi M_{\rm Pl}}{T^6} \frac{n_\gamma(T)}{g_{*,S}(T)\sqrt{g_{*,\rho}(T)}} \langle \Gamma_{\gamma\gamma \to \chi \bar{\chi}} \rangle \,.
    \label{eq:FIAxialvectorcoup}
\end{equation}

\subsection{Pseudoscalar operator} 

For DM interactions with the quark pseudoscalar operator, $\gamma \gamma \to \chi \bar{\chi}$ is the leading process for DM thermalisation and freeze-in production at $T \lesssim 10\,{\rm MeV}$. This process proceeds via the Feynman diagram shown in the left panel of \cref{fig:PSAVcoupFD} and its amplitude can be readily obtained from the leading order Chiral Lagrangian in \cref{eq:DMPScoupChPT}:
\begin{equation}
    i\mathcal{M}_{\gamma\gamma \to \chi \bar{\chi}} = \frac{i \alpha}{\pi \Lambda^2} B_0 \left(C^P_u-C^P_d\right)\, \frac{\epsilon^{\mu\nu\alpha\beta}\,k_{1\mu}k_{2\nu}\epsilon_\alpha(k_1)\epsilon_\beta(k_2)}{s-m^2_{\pi^0}+i m_{\pi^0}\Gamma_{\pi^0}} \bar u_\chi(p_1) \gamma^5 v_\chi(p_2) \,.
    \label{eq:PScoupamp}
\end{equation}
We obtain the corresponding cross-section,
\begin{equation}
    \sigma_{\gamma\gamma \to \chi\bar{\chi}}(s) = \frac{\alpha^2\,}{64\pi^3 \Lambda^4} B^2_0 (C^P_u-C^P_d)^2 \sqrt{1-\frac{4m^2_\chi}{s}} \frac{s^2}{(s-m^2_{\pi^0})^2+m^2_{\pi^0}\Gamma^2_{\pi^0}} \,. 
    \label{eq:PScasexsec}
\end{equation}
Substituting this cross-section into \cref{eq:AVcouprate} yields the thermally-averaged rate, which can then be used in \cref{eq:FIAxialvectorcoup} to obtain the irreducible freeze-in abundance of DM.

\section{Meson decays}
\label{app:flavour}

In this appendix, we provide the meson decay rates used to derive the relevant flavour physics constraints for each type of DM-SM interaction.

\subsection{Vector operator}

For the vector operator, the relevant flavour physics constraints come from the upper-limits on the branching ratios of $K^+ \to \pi^+ \nu\bar{\nu}$~\cite{NA62:2024pjp} and $\pi^0 \to \gamma\nu \bar{\nu}$~\cite{NA62:2019meo}. We provide the decay rates for the corresponding decays to DM, i.e., $K^+ \to \pi^+ \chi\bar\chi$ and $\pi^0 \to \gamma \chi\bar\chi$,  in the following subsections.

\subsubsection{\texorpdfstring{$K^+ \to \pi^+ \chi\bar{\chi}$}{K decay}}

\begin{figure*}[htb!]
    \centering
    \includegraphics[width=0.8\linewidth]{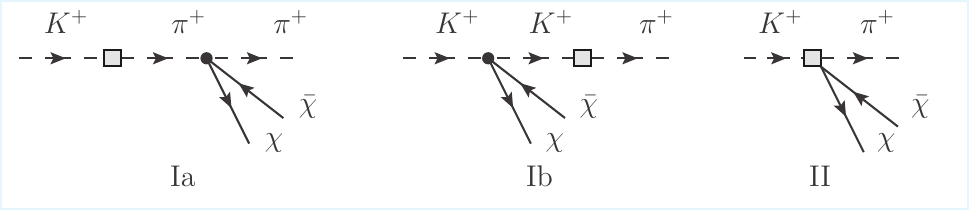}
    \caption{Feynman diagrams contributing to $K^+ \to \pi^+ \chi\bar{\chi}$ for DM interactions with the quark vector current. The grey squares denote the $\Delta S=1$ vertices.} 
    \label{fig:Kaondecayvector}
\end{figure*}

Under the assumption of flavour-diagonal Wilson coefficients $C^V$, the $K \to \pi$ transition is mediated purely by the SM weak interactions. At the quark level, the $s \to d$ transition is mediated by loops containing $W$ bosons; at lower energies, the four-quark operators of the effective weak Lagrangian can be matched onto the Chiral EFT. The leading order $\Delta S = 1$ Lagrangian is then given by \cref{eq:DeltaS1Lagterm1}. 

Expanding \cref{eq:DeltaS1Lagterm1} to quadratic order in the meson fields, the relevant terms are
\begin{align}
    \mathcal{L}^{\rm LO}_{\Delta S=1} \supset & -\sqrt{2}G_F V_{ud}V^*_{us} f^2 \frac{1}{3}\left(3g_8+2g_{27} \right) \,\notag\\ 
    & \bigg[ (\partial_\mu K^+)(\partial^\mu \pi^-) + \frac{i}{\Lambda^2} \left( (C^V_u - C^V_d) \pi^- (\partial_\mu K^+) - (C^V_u - C^V_s) K^+ (\partial_\mu \pi^-) \right) (\bar{\chi}\gamma^\mu \chi) \bigg] + {\rm h.c.} \,,
    \label{eq:DeltaS1transitionvector}
\end{align}
where the term $\propto g^S_8$ identically vanishes.
The first term above, combined with the DM-meson interactions in \cref{eq:DMVcoupChPT}, generates the contributions to the decay $K^+ \to \pi^+ \chi\bar{\chi}$ shown in diagrams Ia and Ib of \cref{fig:Kaondecayvector}. The other terms in \cref{eq:DeltaS1transitionvector} give the contribution in diagram II. Combining all three diagrams, we obtain the amplitude for the $K^+ \to \pi^+ \chi\bar{\chi}$ decay (for fermionic DM),
\begin{equation}
    i\mathcal{M}_{K^+ \to \pi^+ \chi \bar{\chi}} = \frac{i\sqrt{2}}{3\Lambda^2} G_F V_{ud} V^*_{us} f^2 \left(3g_8+2g_{27} \right)  (C^V_d - C^V_s) \left( \frac{m^2_{K^+}+m^2_{\pi^+}}{m^2_{K^+}-m^2_{\pi^+}}\right)\,\bar{u}_\chi(p_1) \cancel{p}_\pi v_\chi(p_2)\,.
    \label{eq:KpiVectoramp}
\end{equation}
The resulting $K^+ \to \pi^+ \chi \bar{\chi}$ decay rate is given by (see also~\cite{DiLuzio:2023xyi})
\begin{equation}
    \Gamma_{K^+ \to \pi^+ \chi \bar{\chi}} = \frac{m^5_{K^+}}{576\pi^3 \Lambda^4}G^2_F |V_{ud}|^2 |V^*_{us}|^2 f^4 \left(3g_8+2g_{27} \right)^2 (C^V_d - C^V_s)^2 \mathcal{F}^V_K(m_{K^+},m_{\pi^+},m_\chi) \,,
    \label{eq:KaondecayVector}
\end{equation}
with
\begin{align}\label{eq:Vector3bodyfun}
    \mathcal{F}^V_K(m_{K^+},m_{\pi^+},m_\chi) &= \int_{\frac{2m_{\pi^+}}{m_{K^+}}}^{1+\frac{m^2_{\pi^+}-4m^2_\chi}{m^2_{K^+}}} dx \int_{y^-(x)}^{y^+(x)} dy \,\, \times \left( \frac{m^2_{K^+}+m^2_{\pi^+}}{m^2_{K^+}-m^2_{\pi^+}}\right)^2
    \begin{cases}
    -2 \left((1-y)(1-y-x)+\frac{m^2_{\pi^+}}{m^2_{K^+}}\right) \quad \text{(fermionic DM)} \,, \\
    \left(1-y-\frac{x}{2} \right)^2 \quad \text{(scalar DM)} \,, \\
    \end{cases}
\end{align}
where $E_{\pi^+} = x m_{K^+}/2$ is the energy of the $\pi^+$, $E_\chi = y m_{K^+}/2$ is the energy of the DM particle, and
\begin{equation}
    y^{\pm}(x) = \frac{1}{2}\bigg[ (2-x) \pm \sqrt{1-\frac{4m^2_\chi}{m^2_{K^+} (1-x) + m^2_{\pi^+}}} \sqrt{x^2-\frac{4m^2_{\pi^+}}{m^2_{K^+}}}\bigg] \,.
    \label{eq:ylims}
\end{equation}
In the limit $m_{K^+} \gg m_{\pi^+},m_\chi$ we obtain $\mathcal{F}^V_K \to 1/12\ (1/48)$ for fermionic (scalar) DM. Notice from \cref{eq:KaondecayVector} that the $K^+ \to \pi^+ \chi \bar{\chi}$ decay rate vanishes when $C^V_d=C^V_s$.

\subsubsection{\texorpdfstring{$\pi^0 \to \gamma\chi \bar{\chi}$}{..}}

When the $K^+ \to \pi^+ \chi \bar{\chi}$ decay rate vanishes (i.e. when $C^V_s=C^V_d$), the leading flavour physics constraint for the vector operator is from the decay $\pi^0 \to \gamma\chi \bar{\chi}$. This process is generated by the WZW functional (see the fourth line of \cref{eq:DMVcoupChPT}), with the relevant Feynman diagram shown on the right in \cref{fig:VcoupMeson}. The corresponding decay rate is given by
\begin{equation}
    \Gamma_{\pi^0 \to \gamma \chi\bar{\chi}} = \frac{\alpha m_{\pi^0}^7}{1024\pi^6\,f^2\,\Lambda^4} (2C^V_u+C^V_d)^2 \mathcal{F}^V_{\pi^0}(m_{\pi^0},m_\chi) \,,
    \label{eq:pi03bodydecays}
\end{equation}
where
\begin{equation}
    \mathcal{F}^V_{\pi^0}(m_{\pi^0},m_\chi)= \int_0^{1-\frac{4m^2_\chi}{m^2_{\pi^0}}} dx \int_{y^-(x)}^{y^+(x)}dy \times 
    \begin{cases}
    \frac{2m^2_\chi}{m^2_{\pi^0}} x^2 + 2(1-x)(1+y^2) - (1-x)(2-x)(x+2y) \quad \text{(fermionic DM)} \,,\\
    -\Big[\frac{m^2_\chi}{m^2_{\pi^0}} x^2 + (1-x)(1-y)(1-x-y)\Big] \quad \text{(scalar DM)} \,,\\
    \end{cases}
\end{equation}
with $E_\gamma = x m_{\pi^0}/2$ the energy of the photon, $E_\chi =  y m_{\pi^0}/2$ the energy of $\chi$ or $\bar\chi$, while
\begin{equation}
    y^{\pm}(x) = \frac{1}{2}\left[ (2-x) \pm x\sqrt{1-\frac{4m^2_\chi}{m^2_{\pi^0}(1-x)}} \,\right] \,.
\end{equation}
In the limit $m_\chi /m_{\pi^0} \to 0$, we find $ \mathcal{F}^V_{\pi^0} \to 1/30\, (1/120)$ for fermionic (scalar) DM.

\subsection{Axial-vector operator}
\label{app:meson-AV}

For the axial-vector operator, the relevant flavour physics constraints come from the measurements of, or upper-limits on, the branching ratios of $K^+ \to \pi^+ \nu\bar{\nu}$~\cite{NA62:2024pjp}, $\pi^0 \to \nu\bar{\nu}$~\cite{NA62:2020pwi}, and $\eta \to \nu\bar{\nu}$~\cite{NA64:2024mah}. In the following subsections we provide the decay rates for the corresponding decays to DM.

\subsubsection{\texorpdfstring{$K^+ \to \pi^+ \chi\bar{\chi}$}{K decay}}

\begin{figure}[t]
    \centering
    \includegraphics[width=0.5\linewidth]{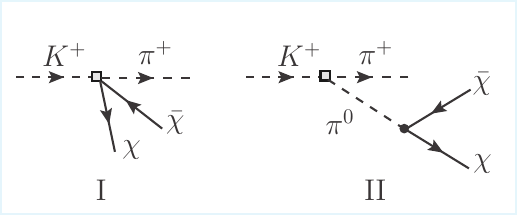}
    \caption{Feynman diagrams contributing to $K^+ \to \pi^+ \chi \bar\chi$ for DM interactions with the quark axial-vector operator. Only Diagram II contributes for interactions with the quark pseudoscalar operator and $G_{a\mu\nu}\tilde{G}^{a\mu\nu}$.} 
    \label{fig:PSAVcoupKaon}
\end{figure}

The $K^+ \to \pi^+ \chi\bar{\chi}$ decay is driven by the $\Delta S = 1$ Lagrangian in \cref{eq:DeltaS1Lagterm1}, which for DM interacting with the quark axial-vector current contains the following relevant terms
\begin{multline}
    \mathcal{L}^{\rm LO}_{\Delta S=1}  \supset -\sqrt{2}G_F V_{ud}V^*_{us}f^2 \Bigg[ \frac{i}{6f} \Big( \left[3g_8+2g_{27} \right] (\partial_\mu K^+) \left[ \pi^- (\partial^\mu \pi^0) -  \pi^0 (\partial^\mu \pi^-)\right] + 5g_{27} (\partial^\mu \pi^0) \left[\pi^- (\partial_\mu K^+) - K^+ (\partial_\mu \pi^-) \right] \Big)\\
    + \frac{i}{\Lambda^2} \bigg( \left[g_8  (C^A_u+C^A_s) + g^S_8 \sum\nolimits_q C^A_q + \frac{g_{27}}{3}(4C^A_u-3C^A_d-C^A_s)\right] \pi^- (\partial_\mu K^+) \\
    - \left[g_8 (C^A_u+C^A_d) +  g^S_8 \sum\nolimits_q C^A_q  + \frac{g_{27}}{3}(4C^A_u-C^A_d-3C^A_s)\right]\, K^+ (\partial_\mu \pi^-) \bigg) (\bar \chi \gamma^\mu \gamma^5 \chi) \Bigg] + {\rm h.c.} \,.
    \label{eq:DeltaS1transitionaxialvector}
\end{multline}
The terms in the second and third lines directly give a contribution to $K^+ \to \pi^+ \chi\bar{\chi}$ via diagram I in \cref{fig:PSAVcoupKaon}. There is a second contribution mediated by the $\pi^0$ (diagram II), which arises from the $K^+ \pi^- \pi^0$ terms in the first line of \cref{eq:DeltaS1transitionaxialvector} combined with the DM coupling to a single $\pi^0$ in \cref{eq:DMAVcoupChPT}. There is a similar contribution mediated by an off-shell $\eta$, which, being suppressed compared to $\pi^0$-mediated diagram, is neglected. 

Combining these two diagrams, we obtain the amplitude for the $K^+ \to \pi^+ \chi\bar{\chi}$ decay as,
\begin{align}
    i\mathcal{M}_{K^+ \to \pi^+ \chi \bar{\chi}} = & -\frac{i\sqrt{2}}{\Lambda^2} G_F V_{ud} V^*_{us} f^2 \Big[ \mathcal{C}^A_{K^+} p_{K^+}  + \mathcal{C}^A_{\pi^+} p_{\pi^+} \Big]_\mu \bar u_\chi (p_1) \gamma^\mu \gamma^5 v_\chi(p_2) \,,
\end{align}
where $\mathcal{C}^A_{K^+}$ and $\mathcal{C}^A_{\pi^+}$ are functions of $q^2_{12}=(p_{K^+}-p_{\pi^+})^2$ given by
\begin{align}
    \mathcal{C}^A_{K^+} &=  \left[ g_8 (C^A_u+C^A_s) - \frac{1}{6} (C^A_u-C^A_d) \frac{ \left(3g_8+ 2g_{27}\right) (q^2_{12}-m^2_{\pi^+}) + 5g_{27} (m^2_{K^+}-m^2_{\pi^+})}{q^2_{12}-m^2_{\pi^0}+ i m_{\pi^0}\Gamma_{\pi^0}} \right] +g^S_8 \sum_q C^A_q + \frac{g_{27}}{3} (4C^A_u-3C^A_d-C^A_s),\\
    \mathcal{C}^A_{\pi^+} &= \left[g_8 (C^A_u+C^A_d) + \frac{1}{6} (C^A_u-C^A_d)  \frac{\left(3g_8+ 2g_{27} \right) (q^2_{12}-m^2_{\pi^+}) + 5g_{27} (m^2_{K^+}-m^2_{\pi^+})}{q^2_{12}-m^2_{\pi^0}+ i m_{\pi^0}\Gamma_{\pi^0}}\right] +g^S_8 \sum_q C^A_q + \frac{g_{27}}{3} (4C^A_u-C^A_d-3C^A_s)\,.
\end{align}
The resulting decay rate is
\begin{equation}
    \Gamma_{K^+ \to \pi^+ \chi \bar{\chi}} = \frac{m^5_{K^+}}{64\pi^3\,\Lambda^4} G^2_F |V_{ud}|^2 |V^*_{us}|^2 f^4 \mathcal{F}^A_K(m_{K^+},m_{\pi^+},m_\chi) \,,
    \label{eq:KaondecayAxialVector}
\end{equation}
with 
\begin{align}
    \mathcal{F}^A_K(m_{K^+},m_{\pi^+},m_\chi) &= -\int_{\frac{2m_{\pi^+}}{m_{K^+}}}^{1+\frac{m^2_{\pi^+}-4m^2_\chi}{m^2_{K^+}}} dx \int_{y^-(x)}^{y^+(x)} dy \, \Bigg[ \left|\mathcal{C}^A_{\pi^+}\right|^2 \left( \frac{m^2_{\pi^+}}{m^2_{K^+}} \left(1- \frac{4 m^2_\chi}{m^2_{K^+}}\right) + (1-y)(1-x-y)\right) \notag\\
    & + 2 {\rm Re}[\mathcal{C}^{A}_{K^+} \mathcal{C}^{A\,*}_{\pi^+}] \left(\frac{m^2_{\pi^+}-2 m^2_\chi x}{m^2_{K^+}} + (1-y)(1-x-y) \right)  + \left|\mathcal{C}^A_{K^+}\right|^2 \left(\frac{m^2_{\pi^+}-4m^2_\chi}{m^2_{K^+}} + (1-y)(1-x-y) \right) \Bigg] \,, \label{eq:KaondecayAxialVectorIntegral} 
\end{align}
where $y^{\pm}(x)$ are given in \cref{eq:ylims}, and $x,y$ have the same definition as in the vector operator case. In the limit $m_{K^+} \gg m_{\pi^+},m_\chi$ we obtain
\begin{equation}
    \mathcal{F}^A_K \to \frac{1}{24} \bigg[\left(g_8 + 4g_{27}\right) C^A_u + \left(g_8 + 2g^S_8 - \frac{4g_{27}}{3} \right) \sum_q C^A_q\bigg]^2 \,.
\end{equation}

There is an important subtlety that arises when using the NA62 measurement of the branching ratio $K^+ \to \pi^+ \nu\bar{\nu}$ to constrain the decay $K^+ \to \pi^+ \chi\bar{\chi}$. This comes from the fact that the amplitudes for the two processes depend differently on the ``missing mass-squared", $m_\text{miss}^2 = (p_{K^+}-p_{\pi^+})^2$. In particular, diagram II in \cref{fig:PSAVcoupKaon} has a resonant enhancement when the $\pi^0$ is on-shell. The NA62 analysis, however, specifically excludes the kinematic region where the $\pi^0$ is on-shell, $ 0.01\,{\rm GeV}^2 < m_\text{miss}^2 < 0.026\,{\rm GeV}^2$, since the decay $K^+ \to \pi^+ \pi^0$ is used for normalisation. When deriving our constraints, we therefore exclude this region of phase-space from the integral in \cref{eq:KaondecayAxialVectorIntegral} when calculating the $K^+ \to \pi^+ \chi\bar{\chi}$ branching ratio. This is expected to result in a conservative bound.

\subsubsection{\texorpdfstring{$\pi^0 \to \chi\bar{\chi}$}{..}}

The $\pi^0 \to \chi\bar{\chi}$ decay proceeds via the Feynman diagram shown on the right of \cref{fig:PSAVcoupFD}, with the coupling from \cref{eq:DMAVcoupChPT}. The corresponding decay rate is given by
\begin{equation}
    \Gamma_{\pi^0 \to \chi\bar{\chi}} = \frac{f^2m^2_\chi m_{\pi^0}}{2\pi \Lambda^4}\sqrt{1-\frac{4m^2_\chi}{m^2_{\pi^0}}} \left[(C^A_u-C^A_d) \bigg(1+\frac{16 B_0 L_4}{f^2} \sum_q m_q \bigg)+ \frac{16 B_0 L_5}{f^2} (C^A_u m_u- C^A_d m_d) \right]^2 \,.
    \label{eq:pi0decayAV}
\end{equation}

\subsubsection{\texorpdfstring{$\eta \to \chi\bar{\chi}$}{..}} 

Similar to the $\pi^0 \to \chi\bar{\chi}$ decay, the $\eta \to \chi\bar{\chi}$ decay proceeds via the right-hand diagram in \cref{fig:PSAVcoupFD}. This decay has the following rate
\begin{equation}
    \Gamma_{\eta \to \chi\bar{\chi}} = \frac{f^2m^2_\chi m_{\eta}}{6\pi \Lambda^4 }\sqrt{1-\frac{4m^2_\chi}{m^2_{\eta}}} \left[ (C^A_u+C^A_d-2C^A_s)\bigg(1+\frac{16 B_0 L_4}{f^2}  \sum_q m_q \bigg) + \frac{16 B_0 L_5}{f^2} (C^A_u m_u+ C^A_d m_d-2 C^A_s m_s)\right]^2 \,.
    \label{eq:etadecayAV}
\end{equation}

\subsection{Pseudoscalar operator} 

Similar to the axial-vector operator, for the pseudoscalar operator, the leading order Chiral Lagrangian \eqref{eq:DMPScoupChPT} leads to the decays $\pi^0 \to \chi \bar{\chi}$ and $\eta \to \chi \bar{\chi}$. There are also constraints from the decay $K^+ \to \pi^+ \chi \bar\chi$.

\subsubsection{\texorpdfstring{$K^+ \to \pi^+ \chi\bar{\chi}$}{K decay}}

The relevant terms in the $\Delta S = 1$ Lagrangian \eqref{eq:DeltaS1Lagterm1} for the pseudoscalar operator are
\begin{align}
    \mathcal{L}^{\rm LO}_{\Delta S = 1} \supset -\sqrt{2}G_F V_{ud}V^*_{us} f^2\,\frac{i}{6f}  & \Big[\left(3g_8+2g_{27}\right) (\partial_\mu K^+) \left[ \pi^- (\partial^\mu \pi^0) -  \pi^0 (\partial^\mu \pi^-) \right] + 5g_{27} (\partial^\mu \pi^0) \left[ \pi^- (\partial_\mu K^+) - K^+ (\partial_\mu \pi^-) \right] \Big] + {\rm h.c.} \,.
    \label{eq:DeltaS1transitionpseudoscalar}
\end{align}
The $K^+ \to \pi^+ \chi\bar\chi$ decay then proceeds via an intermediate $\pi^0$ (diagram II in \cref{fig:PSAVcoupKaon}), with the amplitude given by
\begin{equation}
    i\mathcal{M}_{K^+ \to \pi^+ \chi \bar{\chi}} =  -\frac{i}{3\sqrt{2}\Lambda^2} B_0 (C^P_u-C^P_d) G_F V_{ud} V^*_{us} f^2 \frac{\left(3g_8+ 2g_{27}\right) (q^2_{12}-m^2_{\pi^+}) + 5g_{27} (m^2_{K^+}-m^2_{\pi^+})}{q^2_{12}-m^2_{\pi^0}+ i m_{\pi^0}\Gamma_{\pi^0}} \bar u_\chi (p_1) \gamma^5 v_\chi(p_2) \,,
\end{equation}
where $q^2_{12} = (p_{K^+}-p_{\pi^+})^2$. 
From this amplitude we obtain the decay rate
\begin{align}
    \Gamma_{K^+ \to \pi^+ \chi\bar{\chi}} &= \frac{m_{K^+}}{2304\pi^3 \Lambda^4} G^2_F |V_{ud}|^2 |V^*_{us}|^2 f^4 B^2_0 (C^P_u-C^P_d)^2 \mathcal{F}^P_K(m_{K^+},m_{\pi^+},m_\chi) \,,
    \label{eq:KaondecayPseudoscalar}
\end{align}
with
\begin{align}
    \mathcal{F}^P_K(m_{K^+},m_{\pi^+},m_\chi) &= \int_{\frac{2m_{\pi^+}}{m_{K^+}}}^{1+\frac{m^2_{\pi^+}-4m^2_\chi}{m^2_{K^+}}} dx \,\bigg[ \left(3g_8+2g_{27} \right) m^2_{K^+}(1-x) + 5g_{27} (m^2_{K^+}-m^2_{\pi^+}) \bigg]^2 \times \notag\\
    & \frac{\left(m^2_{\pi^+}+m^2_{K^+}(1-x)\right)}{\left(m^2_{\pi^+} - m^2_{\pi^0} +m^2_{K^+}(1-x)\right)^2 + m^2_{\pi^0}\Gamma^2_{\pi^0}} \sqrt{1-\frac{4m^2_\chi}{m^2_{\pi^+} + m^2_{K^+} (1-x) }} \sqrt{x^2-\frac{4m^2_{\pi^+}}{m^2_{K^+}}} \,,
    \label{eq:KaondecayPseudoscalarsimple}
\end{align}
where $x$ has the same definition as previously for the vector and axial-vector operators.

As discussed for the axial-vector operator in \cref{app:meson-AV}, we exclude the region $ 0.01\,{\rm GeV}^2 \leq (p_{K^+}-p_{\pi^+})^2 \leq 0.026\,{\rm GeV}^2$, where the $\pi^0$ is on-shell, from the phase-space integral when imposing the constraint from NA62.

\subsubsection{\texorpdfstring{$\pi^0 \to \chi\bar{\chi}$}{..}} 

The decay rate for $\pi^0 \to \chi\bar{\chi}$ is readily obtained from the Lagrangian in \cref{eq:DMPScoupChPT} and is given by
\begin{equation}
    \Gamma_{\pi^0 \to \chi\bar{\chi}} = \frac{B^2_0 f^2}{8\pi \Lambda^4} (C^P_u-C^P_d)^2 m_{\pi^0}\sqrt{1-\frac{4m^2_\chi}{m^2_{\pi^0}}} \,.
    \label{eq:pi0decayPS}
\end{equation}

\subsubsection{\texorpdfstring{$\eta \to \chi\bar{\chi}$}{..}} 

The decay rate for $\eta \to \chi\bar{\chi}$ (see \cref{eq:DMPScoupChPT}) is 
\begin{equation}
    \Gamma_{\eta \to \chi\bar{\chi}} = \frac{B^2_0 f^2}{24\pi \Lambda^4} (C^P_u+C^P_d-2C^P_s)^2 m_\eta\sqrt{1-\frac{4m^2_\chi}{m^2_\eta}} \,.
    \label{eq:etadecayPS}
\end{equation}

\twocolumngrid

\bibliographystyle{apsrev4-2}
\bibliography{paper}


\end{document}